\documentclass{article}
\usepackage{geometry}
\usepackage[sc,normalsize]{caption2}
\usepackage{amsmath}
\usepackage{graphicx}
\usepackage{ulem}
\usepackage{amsfonts}
\usepackage{amssymb}

\newtheorem{theorem}{Theorem}

\newtheorem{definition}[theorem]{Definition}

\newenvironment{proof}[1][Proof]{\textbf{#1.} }{\ \rule{0.5em}{0.5em}}
\input{tcilatex}

\begin{document}

\title{Geometric Approach to Digital Quantum Information}
\author{Chad Rigetti$^{1}$, R\'{e}my Mosseri$^{2}$ and Michel Devoret$^{1}$ 
\\
$^{1}$Department of Applied Physics, Yale University, \\
New Haven, Connecticut 06520-8284, USA\\
$^{2}$Groupe de Physique des Solides, Universit\'{e} Paris VI,\\
campus Boucicaut, 140 rue de Lourmel, 75015 Paris, France}
\maketitle

\begin{abstract}
We present geometric methods for uniformly discretizing the continuous $N$%
-qubit Hilbert space $H_{N}$. When considered as the vertices of a
geometrical figure, the resulting states form the equivalent of a Platonic
solid. The discretization technique inherently describes a class of $\pi/2$
rotations that connect neighboring states in the set, i.e. that leave the
geometrical figures invariant. These\ rotations are shown to generate the
Clifford group, a general group of discrete transformations on $N$ qubits.
Discretizing $H_{N}$ allows us to define its \textit{digital} \textit{%
quantum information} content, and we show that this information content
grows as $N^{2}$. While we believe the discrete sets are interesting because
they allow extra-classical behavior---such as quantum entanglement and
quantum parallelism---to be explored while circumventing the continuity of
Hilbert space, we also show how they may be a useful tool for problems in
traditional quantum computation. We describe in detail the discrete sets for
one and two qubits.
\end{abstract}

\section{Introduction}

The discrete nature of the configuration space for $N$ classical bits is the
key property allowing robustness of digital computation. The Hilbert space $%
H_{N}$ for $N$ qubits, on the other hand, is a \textit{continuous} complex%
\textbf{\ }manifold. This continuity appears essential to the exponential
speed-up of some quantum computing algorithms, such as Shor's factoring
algorithm\cite{Shor}, over their classical counterparts\footnote{%
As evinced by the Gottesman-Knill theorem, for example. See \cite{Got-Heis}.}%
. But it also poses a challenging problem for the experimentalist: errors in
quantum gates are themselves continuous, so even minute errors can
accumulate throughout the execution of an algorithm and lead to its failure.

Yet, quantum error correction and fault-tolerant computation schemes have
been developed to meet this challenge\cite{Got96,Steane,Shor95}. That
reliable quantum computation is possible using both a noisy quantum register
and noisy gates is a result of surpassing importance. However, such schemes
still place stringent fidelity requirements on the basic quantum gates and
the quantum register: estimates for the threshold error probability above
which they fail are typically $10^{-5}$--$10^{-6}$\cite{Nielsen-Chuang}.

Can universal control of a scalable quantum register with this level of
fidelity be realized? If so, are there concepts we can borrow from digital
computation that might facilitate the development of this technology? If
not, are there intermediate computational paradigms that might relax these
requirements, but still exploit \textquotedblleft
extra-classical\textquotedblright\ phenomena such as quantum parallelism and
quantum entanglement? We note that these are still possible in a \textit{%
discrete }Hilbert space.

Quantum gates are typically implemented by applying time-dependent fields to
a qubit system. They correspond to rotations of a unit vector in $H_{N}$,
with the angle of rotation usually determined by the\ duration and amplitude
of the pulse which generates the field. In principle, such rotations are
simple to implement, given an appropriate time-dependent Hamiltonian. But in
practice, noise in both the qubit system and applied fields inevitably leads
to errors. Sophisticated techniques that build up a desired gate from a
sequence of rotations about successively orthogonal axes have been developed
to mitigate the effects of noise. In the field of NMR, especially,
techniques for performing high-fidelity rotations are now very mature\cite%
{Cummins}. Yet such techniques for protecting against noise are not directly
generalizable to arbitrary angles and axes of rotation. As a result, most
experimental protocols for quantum manipulations rely as heavily as possible
on a small set of rotations, usually by an angle of $\pi/2$ or $\pi$,
specifically optimized for the given qubit system. In the landmark
experiment by Vandersypen, et al., in which an NMR-based quantum processor
was used to factor the number fifteen via Shor's algorithm, the protocol
contained a single rotation by an angle less than $\pi/2$---a conditional $%
\pi/4$ rotation\cite{Vandersypen}.

Nonetheless, universal control of a quantum register requires in theory only
a finite number of discrete gates, provided the gates form a universal set.
Then an arbitrary \textquotedblleft software-level\textquotedblright\
quantum gate can be constructed to a precision $\epsilon$ by concatenating $%
\mathcal{O}(\log^{c}(1/\epsilon))$ discrete gates from the universal set ($%
c\approx 2$)\cite{Nielsen-Chuang}. However, given that each discrete gate
itself would likely comprise a sophisticated series of rotations, the
prospect of concatenating $\mathcal{O}(\log^{c}(1/\epsilon))$ such gates to
create each software-level operation---and doing so before the register
decoheres---makes the fidelity requirements of fault-tolerant computing
schemes all the more exacting.

Much of this difficulty in achieving high-fidelity control of a quantum
register can be alleviated by limiting ourselves to \textit{non}universal
sets of quantum gates which generate only \textit{finite} transformation
groups. A finite transformation group implies a finite number of possible
states, so this is equivalent to imposing a discretization on the underlying
Hilbert space: the quantum register becomes \textquotedblleft
digital\textquotedblright. By suitably choosing the transformation group,
the allowed states can be selected to have certain well-defined properties,
such as known expectation values with respect to a set of measurement
operators.

As an illustration of this idea, consider the task of testing the
experimental protocol for generating one-qubit rotations, which can be
represented on the Bloch sphere. Suppose we wish to optimize the fidelity of
a $\pi/2$ rotation about the $y$-axis in a given qubit system. Starting from
the state $\left\vert 0\right\rangle $, we perform a counter-clockwise $%
\pi/2 $ rotation about the $y$-axis, yielding the target state $\left\vert
+x\right\rangle =(\left\vert 0\right\rangle +\left\vert 1\right\rangle )/%
\sqrt{2}$, then we perform a measurement in the $\{\left\vert 0\right\rangle
,\left\vert 1\right\rangle \}$ basis. By repeating this many times, we
obtain the expectation value $\left\langle \sigma_{z}\right\rangle $ of the
target state $\left\vert +x\right\rangle $. Assuming imperfections in the
state preparation and readout have been accounted for, this expectation
value would approach zero if our $\pi/2$ gate were perfect, since $%
\left\langle +x|\sigma _{z}|+x\right\rangle =0$, while deviations from zero
would imply an imperfect $\pi/2$ gate. Specifically, $\left\langle
\sigma_{z}\right\rangle =\delta$ would imply that, on average, the gate has
performed a rotation by an angle of $2\arccos\sqrt{(\delta+1)/2}$. With
knowledge of other one-qubit discrete states and their expectation values $%
\left\langle \sigma_{i}\right\rangle $, we could also test rotations about
the $x$- and $z$-axes. By direct generalization, this simple protocol can be
used to test rotations on any number of qubits, provided we have an
appropriate discrete set of target states.

Also, in the nascent field of quantum feedback control, techniques have been
developed to \textit{dynamically} correct quantum processes. By performing
continuous weak measurements on the quantum system, it is possible to
control and correct quantum state evolution through feedback\cite{Doherty,
Doherty 2}. Incorporating these techniques into quantum computing
experiments could also be facilitated if the number of processes involved
were reduced to include only a small class of rotations connecting states
with well-defined properties.

Having explained why we wish to consider a discrete subset of the full
continuous Hilbert space, we would now like to draw a geometrical analogy.
Since we do not want to privilege any region of Hilbert space over any
other, the set must comprise a uniform sampling of\textbf{\ }$H_{N}$. The
structure of the finite sets we have in mind is exemplified in real space by
the Platonic solids---geometrical figures such as the tetrahedron, cube and
octahedron characterized by the geometric equivalence of their
vertices---which represent discrete subsets uniformly spanning a sphere in $%
\mathbb{R}
^{3}$. In short, we are seeking to generalize the Platonic solids to Hilbert
space by selecting from $H_{N}$ a finite set of states corresponding to the
vertices of\ a $2^{N}$-dimensional complex uniform polytope. We call such\
subsets\ \textit{uniform Hilbertian polytopes,} and denote them by $%
\mathfrak{H}_{N}$.

In discretizing an $N$-qubit register, what extra-classical phenomena must
be sacrificed? If we select the transformation group to include only $\pi$
rotations---the quantum generalizations of the \textsc{NOT} gate---we
generate only a discrete set of $2^{N}$ states, and fall back on a purely
classical register, with no possibility for quantum entanglement. But, as we
will show, the next level towards finer rotations, the transformation group
based on $\pi/2$ rotations, is sufficient for rich extra-classical behavior:
the number of discrete states in the set then grows as $2^{(N^{2}+3N)/2}$%
\cite{Gottesman Pers}, the majority of which are entangled for $N>2$. Also,
the super-extensive growth of the discrete set relative to the classical
number of states $2^{N}$ implies that a great deal of the quantum
parallelism possible in the full Hilbert space remains possible in the
discrete set. Though such a digital quantum register would not allow
algorithms which are exponentially faster than their classical counterparts
(Gottesman-Knill theorem), a possible reduction of an algorithmic scaling
speed from $O(N^{2})$ to $O(N)$ could still be useful.

At the same time, within the framework of traditional quantum computation, a
discrete set and its associated transformation group can provide a useful
\textquotedblleft roadmap\textquotedblright\ for navigation in the entire
Hilbert space.

The notion of discrete sets of $N$-qubit states is not novel. Indeed,
discrete sets have already been considered in quantum error-correcting
codes. There,\ a special set of $2^{k}$ orthogonal states, to be used as
codewords for the basis states of $k$ encoded qubits, are selected from a
higher-dimensional continuous space $H_{N}$. Gottesman's stabilizer
formalism provides a general framework for describing and producing quantum
error-correcting codes, and allows an analysis of a broad class of quantum
networks in the Heisenberg picture\cite{Got-Heis,Got-Thesis}. Powerful
though it is, the stabilizer formalism approaches the problem of
discretization algebraically; it does not address the geometric relationship
between the discrete quantum states, nor the relationships among the various
gates that connect these states.\ 

The purpose of this paper is thus to provide such a\ geometric approach to
the uniform discretization of $H_{N}$, and to suggest the use of such
discrete sets, either as an arena for exploring extra-classical behavior, or
as a heuristic tool for the analysis of certain quantum information
processing problems. We refer to these notions collectively as \textit{%
digital quantum information}\footnote{%
Later, we will use the phrase rigorously, in reference to the information
content of the discrete Hilbert space.}.\textit{\ }

For simplicity, we focus here on the one- and two-qubit Hilbert spaces.
However,\ most of our results are directly generalizable to
higher-dimensional spaces. When possible, we use a language that makes this
generalization straightforward, if tedious. In section 2, we treat the
discretization problem using stabilizer theory and derive a class of
generalized $\pi/2$ rotations belonging to the Clifford group that connect
states in the discrete set. Later in section 3, we present an alternate,
purely geometric approach to discretization based on shelling the high
dimensional lattices.

\section{Discretization based on stabilizer theory}

\subsection{Stabilizers and the generalized Pauli matrices}

We begin this section with some essential results from stabilizer theory.
First, define the $N$-qubit Pauli group $\mathcal{G}_{N}$ as the set of all $%
N$-fold tensor products of $2\times2$ Pauli matrices, with four possible
overall phases to satisfy the closure requirement:%
\begin{equation*}
\mathcal{G}_{N}=\left\{ \sigma_{w},\sigma_{x},\sigma_{y},\sigma_{z}\right\}
^{\otimes N}\otimes\left\{ \pm1,\pm i\right\} ,
\end{equation*}
where\footnote{%
The present numbering scheme has been chosen to coincide with the binary
vector space representation of stabilizer codes, as in \cite{Got-Thesis}.}%
\begin{align*}
& 
\begin{tabular}{ll}
$\sigma_{w}=\sigma_{0}=\left[ 
\begin{array}{cc}
1 & 0 \\ 
0 & 1%
\end{array}
\right] ,$ & $\sigma_{z}=\sigma_{1}=\left[ 
\begin{array}{cc}
1 & 0 \\ 
0 & -1%
\end{array}
\right] ,$%
\end{tabular}
\ \ \ \ \ \  \\
& 
\begin{tabular}{ll}
$\sigma_{x}=\sigma_{2}=\left[ 
\begin{array}{cc}
0 & 1 \\ 
1 & 0%
\end{array}
\right] ,$ & $\sigma_{y}=\sigma_{3}=\left[ 
\begin{array}{cc}
0 & -i \\ 
i & 0%
\end{array}
\right] .$%
\end{tabular}%
\end{align*}
Clearly, each element of $\mathcal{G}_{N}$ acts on the $N\,$-qubit Hilbert
space. $\mathcal{G}_{N}$ has order $4^{N+1}$, and is generated by a minimal
set of $2N$ elements, i.e. two non-identity $\sigma^{\prime}s$ acting on
each qubit. We refer to individual elements of $\mathcal{G}_{N}$ as \textit{%
generalized} Pauli matrices, and denote them $\Sigma_{\alpha
\beta...\zeta}=\sigma_{\alpha}\otimes\sigma_{\beta}\otimes\cdots\otimes
\sigma_{\zeta}$. The generalized Pauli matrices share many of the properties
of the $2\times2$ Pauli matrices. For example, they all either commute or
anti-commute, and 
\begin{align*}
\Sigma_{j}^{\dagger} & =\Sigma_{j}\text{ (Hermitian),} \\
\Sigma_{j}^{2} & =\mathrm{id}\text{ (Square root of unity),} \\
\mathrm{Tr}\Sigma_{j}^{\dagger}\Sigma_{k} & =2^{N}\delta_{jk}\text{
(Orthogonal).}
\end{align*}

A stabilizer is an Abelian subgroup of the Pauli group. In the present work,
we are predominantly concerned with the commutation properties of the
generalized Pauli matrices, so we neglect the phases $\left\{ \pm1,\pm
i\right\} $ required for closure of $\mathcal{G}_{N}$ under multiplication.
That is, we deal with the \textit{set }$\mathcal{S}_{N}$ of $4^{N}$
generalized Pauli matrices rather than the group $\mathcal{G}_{N}$. To
distinguish the Abelian subsets of $\mathcal{S}_{N}$ from the Abelian
subgroups of $\mathcal{G}_{N}$, we refer to the former as \textit{pseudo}%
stabilizers, a name which also highlights the close relationship between
this work and stabilizer theory.

The largest possible subsets of $\mathcal{S}_{N}$ whose elements all
mutually commute have $2^{N}$ elements. These \textit{maximal}
pseudostabilizers will form the foundation of our first discretization
procedure.

\subsection{\protect\bigskip The uniform Hilbertian polytope $\mathfrak{H}%
_{N}$}

We are now in a position to discuss a formal definition for the uniform
Hilbertian polytope\ for $N$ qubits. First, we establish the desired
properties the discrete sets must have. We seek to construct $\mathfrak{H}%
_{N}$, such that:

1.\qquad It contains all the states $\left\vert
b_{0}b_{1}...b_{N-1}\right\rangle $ corresponding to the classical bit
configurations.

2.\qquad Each state of $\mathfrak{H}_{N}$ is geometrically equivalent to all
the others (uniformity).

3.\qquad The distance between two normalized states $\Psi_{j}$ and $\Psi_{k}$%
, defined as%
\begin{equation*}
d_{jk}=2\cos^{-1}(\left\langle \Psi_{j}|\Psi_{k}\right\rangle )
\end{equation*}
satisfies\footnote{%
Discretizations with a finer minimum distance may be useful and would be
interesting to explore (for two qubits, see section 3). \ For one qubit this
could correspond, for instance, to the icosahedral geometry.}%
\begin{equation*}
d_{jk}\geq\pi/2\mathit{\ for\ all\ j,k}\text{.}
\end{equation*}

4.\qquad It is the largest set of states which satisfies the above
requirements.

Denote by $s_{N}^{a}$ the maximal pseudostabilizers in $\mathcal{S}_{N}$.
Then these desired properties are obtained if we adopt the following
construction for the vertices of $\mathfrak{H}_{N}$:

\begin{definition}
An $N$-qubit state vector is an element of $\mathfrak{H}_{N}$ if and only if
it is a common eigenvector of each element of a \textit{maximal}
pseudostabilizer $s_{N}^{a}$. That is, if $\Sigma_{j}$ is a generalized
Pauli matrix on $N$ qubits belonging to $s_{N}^{a},$ $\left\vert
\Psi_{j}\right\rangle $ is an $N$-qubit state vector, and $\lambda_{j}$ is
an eigenvalue of $\Sigma_{j}$ belonging to the vector $\left\vert \Psi
_{j}\right\rangle ,$%
\begin{equation*}
\left\vert \Psi_{j}\right\rangle \in\mathfrak{H}_{N}\;\mathrm{%
\Leftrightarrow }\;\Sigma_{j}\left\vert \Psi_{j}\right\rangle
=\lambda_{j}\left\vert \Psi _{j}\right\rangle \;for\ all\ \Sigma_{j}\in
s_{N}^{a}.
\end{equation*}
\end{definition}

As a consequence of this definition, and from the theory of stabilizers, we
find:

\qquad a) Each $s_{N}^{a}$, which has $2^{N}-1$ elements different from the
identity, generates $2^{N}$ different discrete states all separated by $%
d_{jk}=\pi$. Each state corresponds to a unique pattern of $\lambda_{j}=\pm1$%
.

\qquad b) Each $s_{N}^{a}$ shares exactly half, or $2^{N-1}$, of its
elements with its nearest neighbors; $2^{N-2}$ with its second-nearest
neighbors, etc. Any discrete state in $\mathfrak{H}_{N}$ therefore has $N$
\textquotedblleft levels\textquotedblright\ of non-orthogonal neighboring
states.

\qquad c) For each $s_{N}^{a}$ and each of its nearest neighbors $s_{N}^{b}$
one can associate by a general algorithm a transformation from the common
eigenvectors of $s_{N}^{a}$ to those of $s_{N}^{b}.$ That is, any two states
of $\mathfrak{H}_{N}$ are linked by a finite sequence of similarity
transformations.

\qquad d) The similarity transformations are formed from generalized
orthogonal $\pi/2$ rotations of the form:

\begin{equation*}
X_{k\,l}^{a}=\frac{1}{\sqrt{2}}\left( \Sigma_{k}+i\Sigma_{l}\right) \text{,}%
\;\ \text{where\ \ }\Sigma_{k},\Sigma_{l}\in s_{N}^{a}\text{.}
\end{equation*}
The superscript $a$ denotes a subset $s_{N}^{a}$ to which both its $%
\Sigma^{\prime}s$ belong and the subscripts specify the $\Sigma^{\prime}s$.
\ The inverse operations are: 
\begin{equation*}
(X_{k\,l}^{a})^{-1}=\frac{1}{\sqrt{2}}\left( \Sigma_{k}-i\Sigma_{l}\right)
=-iX_{l\,k}^{a}\text{.}
\end{equation*}
This definition implies that for any $X$,

\begin{align*}
X^{\dagger}X & =\mathrm{id}\text{ \ (Unitary),} \\
X^{4} & =-\mathrm{id}\text{ \ (}\pi/2\text{ Rotations),}
\end{align*}
which is consistent with the property that a spin-$1/2$ acquires an overall
phase of $e^{i\pi}=-1$ when rotated by $2\pi.$

\qquad e) The $X$'s generate the Clifford group $\mathcal{C}_{N}$, defined
as the normalizer of the Pauli group\cite{Got-Thesis}, which has the
property of leaving $\mathfrak{H}_{N}$\ invariant (proof to follow).

\qquad f) The set $\mathcal{S}_{N}$ of generalized Pauli matrices on $N$
qubits contains%
\begin{equation*}
s=\prod_{k=0}^{N-1}(2^{N-k}+1)
\end{equation*}
maximal pseudostabilizers $s_{N}^{a}$. \ Each has $2^{N}$ elements, and
contributes $2^{N}$ simultaneous eigenvectors. $\ $The uniform Hilbertian
polytope on $N$\ qubits $\mathfrak{H}_{N}$ therefore contains 
\begin{equation*}
V_{N}=2^{N}\prod_{k=0}^{N-1}(2^{N-k}+1)
\end{equation*}
vertices, or states\cite{Gottesman Pers}. \ The following table gives the
first values of $V_{N}$, along with the number of classical bit
configurations for comparison.

\begin{equation*}
\begin{tabular}{|c|c|c|c|c|c|l|l|}
\hline
$N$ & $1$ & $2$ & $3$ & $4$ & $5$ & $6$ & $7$ \\ \hline\hline
$V_{N}$ & $6$ & $60$ & $1080$ & $36,720$ & $2,423,520$ & $315,057,600$ & $%
81,284,860,800$ \\ \hline
\multicolumn{1}{|l|}{$C_{N}$} & \multicolumn{1}{|l|}{$2$} & 
\multicolumn{1}{|l|}{$4$} & \multicolumn{1}{|l|}{$8$} & \multicolumn{1}{|l|}{%
$16$} & \multicolumn{1}{|l|}{$32$} & $64$ & $128$ \\ \hline
\end{tabular}
\ 
\end{equation*}
The digital quantum information in \thinspace$N$ qubits can be defined as
the information content of $\mathfrak{H}_{N}$, i.e. as $\log_{2}V_{N}$. It
is easy to show that $V_{N}$ grows as $2^{(N^{2}+3N)/2}$, so this
information content is super-extensive in $N$. While it is insufficient for
algorithms which would be exponentially faster than classical ones, it is
nonetheless a remarkable property for a discrete space.

We now turn to an explicit construction of the uniform Hilbertian polytope
for the one- and two-qubit cases.

\subsection{The one-qubit case, $\mathfrak{H}_{1}$}

We show here that $\mathfrak{H}_{1}$ is isomorphic to an octahedron. For one
qubit, the set $\mathcal{S}_{N}$ is simply the Pauli matrices: $\mathcal{S}%
_{1}=\left\{ \sigma_{w},\sigma_{x},\sigma_{y},\sigma_{z}\right\} $. The last
three $\sigma^{\prime}s$ anti-commute with one another, while they all
commute with the identity $\sigma_{w}$. So the three sets of mutually
commuting matrices are trivial to construct: $s_{1}^{1}=\left\{
\sigma_{w},\sigma _{z}\right\} ,$ $s_{1}^{2}=\left\{
\sigma_{w},\sigma_{x}\right\} $ and $s_{1}^{3}=\left\{
\sigma_{w},\sigma_{y}\right\} .$

When the elements of $s_{1}^{1}$ are diagonalized, we obtain the
computational basis: 
\begin{align*}
\left\vert +z\right\rangle & =\left\vert 0\right\rangle , \\
\left\vert -z\right\rangle & =\left\vert 1\right\rangle .
\end{align*}
$s_{1}^{2}$ generates the pair

\begin{align*}
\left\vert +x\right\rangle & =\frac{\left\vert 0\right\rangle +\left\vert
1\right\rangle }{\sqrt{2}}, \\
\left\vert -x\right\rangle & =\frac{\left\vert 0\right\rangle -\left\vert
1\right\rangle }{\sqrt{2}}.
\end{align*}
And $s_{1}^{3}$ generates

\begin{align*}
\left\vert +y\right\rangle & =\frac{\left\vert 0\right\rangle +i\left\vert
1\right\rangle }{\sqrt{2}}, \\
\left\vert -y\right\rangle & =\frac{\left\vert 0\right\rangle -i\left\vert
1\right\rangle }{\sqrt{2}}.
\end{align*}

There are three orthogonal $\pi/2$ rotations, which form the \textit{seed}
of $\mathfrak{H}_{1}$:%
\begin{equation*}
\begin{tabular}{l}
$X_{0\,1}^{1}=\frac{1}{\sqrt{2}}\left( \sigma_{0}+i\sigma_{1}\right) ,$ \\ 
$X_{0\,2}^{2}=\frac{1}{\sqrt{2}}\left( \sigma_{0}+i\sigma_{2}\right) ,$ \\ 
$X_{0\,3}^{3}=\frac{1}{\sqrt{2}}\left( \sigma_{0}+i\sigma_{3}\right) .$%
\end{tabular}
\ 
\end{equation*}
The diagonalization of the seed elements leads directly to the six
eigenstates, as summarized in the table below. The states are listed here as
unnormalized row vectors for clarity, and are separated into columns
according to their eigenvalues. 
\begin{equation*}
\begin{tabular}{|l|l|l|}
\hline
Set & $1+i$ & $1-i$ \\ \hline
$1$ & $(1,0)$ & $(0,1)$ \\ \hline
$2$ & $(1,1)$ & $(1,-1)$ \\ \hline
$3$ & $(1,i)$ & $(1,-i)$ \\ \hline
\end{tabular}
\ 
\end{equation*}
\qquad

Each of the $\pi/2$ rotations has an inverse:

\begin{equation*}
(X_{0\,k}^{k})^{-1}=-iX_{k\,\,0}^{k}=\frac{1}{\sqrt{2}}\left( \sigma
_{0}-i\sigma_{k}\right) \text{.}
\end{equation*}
It is easy to verify that 
\begin{equation*}
(X_{0\,k}^{k})^{2}=\frac{1}{2}\left( \sigma_{0}+i\sigma_{k}\right)
^{2}=i\sigma_{k}
\end{equation*}
and that the $X^{\prime}s$ are mapped into one another by similarity
transformation: 
\begin{align*}
X_{0\,\,j}^{j}X_{0\,i}^{i}\left( X_{0\,j}^{j}\right) ^{-1} & =\frac {1}{2%
\sqrt{2}}\left( \sigma_{0}+i\sigma_{j}\right) \left( \sigma
_{0}+i\sigma_{i}\right) \left( \sigma_{0}-i\sigma_{j}\right) \\
& =\frac{1}{2\sqrt{2}}\left( 2\sigma_{0}+2i\text{\ }\epsilon_{ijk}\sigma
_{k}\right) \\
& =X_{0\,k}^{k}\text{ \ \ \ if \ \ \ }\epsilon_{ijk}=1 \\
& =\text{ }\left( X_{0\,k}^{k}\right) ^{-1}\text{\ \ \ if \ \ \ }%
\epsilon_{ijk}=-1
\end{align*}
This implies that each $X$ transforms a member of $\mathfrak{H}_{1}$ into
its neighbor.

\begin{proof}
\textit{If } 
\begin{equation*}
X^{j}\left\vert \Psi_{j}\right\rangle =\lambda_{j}\left\vert \Psi
_{j}\right\rangle
\end{equation*}
\textit{and if} 
\begin{equation*}
\left\vert \Psi_{i(j)}\right\rangle =X^{i}\left\vert \Psi_{j}\right\rangle ,
\end{equation*}
\textit{\ \ then } 
\begin{align*}
X^{i}X^{j}(X^{i})^{-1}\left\vert \Psi_{i(j)}\right\rangle &
=X^{i}X^{j}\left\vert \Psi_{j}\right\rangle \\
& =\lambda_{j}X^{i}\left\vert \Psi_{j}\right\rangle \\
& =\lambda_{j}\left\vert \Psi_{i(j)}\right\rangle
\end{align*}
\textit{therefore }$\left\vert \Psi_{i(j)}\right\rangle $\textit{\ is an
eigenvector of }$X^{k}$.
\end{proof}

The $X^{\prime}s$ with their inverse generate a twenty-four element group
isomorphic to the octahedral group of pure rotations which leaves the
octahedron invariant.

\subsection{The two-qubit case, $\mathfrak{H}_{2}$}

The set of generalized Pauli matrices for two qubits $\mathcal{S}_{2}$
comprises $4^{2}=16$,\ $2^{2}\times2^{2}$ matrices given by $\Sigma
_{\lambda\mu}=\sigma_{\lambda}\otimes\sigma_{\mu},$ $\lambda,\mu=w,x,y,z$,
as presented below. We write the index on the $\sigma^{\prime}s$ in binary,
then concatenate the two strings to form the new index for the $%
\Sigma^{\prime}s$. For example, $\sigma_{y}\otimes\sigma_{z}=\sigma_{3}%
\otimes\sigma_{1}=\sigma_{11}\otimes\sigma_{01}=\Sigma_{1101}=\Sigma_{13}.$

\begin{equation*}
\begin{tabular}{cc}
$\Sigma_{ww}=\Sigma_{0}=\left[ 
\begin{array}{cccc}
1 & 0 & 0 & 0 \\ 
0 & 1 & 0 & 0 \\ 
0 & 0 & 1 & 0 \\ 
0 & 0 & 0 & 1%
\end{array}
\right] $ & $\Sigma_{wz}=\Sigma_{1}=\left[ 
\begin{array}{cccc}
1 & 0 & 0 & 0 \\ 
0 & -1 & 0 & 0 \\ 
0 & 0 & 1 & 0 \\ 
0 & 0 & 0 & -1%
\end{array}
\right] $ \\ 
$\Sigma_{wx}=\Sigma_{2}=\left[ 
\begin{array}{cccc}
0 & 1 & 0 & 0 \\ 
1 & 0 & 0 & 0 \\ 
0 & 0 & 0 & 1 \\ 
0 & 0 & 1 & 0%
\end{array}
\right] $ & $\Sigma_{wy}=\Sigma_{3}=\left[ 
\begin{array}{cccc}
0 & -i & 0 & 0 \\ 
i & 0 & 0 & 0 \\ 
0 & 0 & 0 & -i \\ 
0 & 0 & i & 0%
\end{array}
\right] $ \\ 
$\Sigma_{zw}=\Sigma_{4}=\left[ 
\begin{array}{cccc}
1 & 0 & 0 & 0 \\ 
0 & 1 & 0 & 0 \\ 
0 & 0 & -1 & 0 \\ 
0 & 0 & 0 & -1%
\end{array}
\right] $ & $\Sigma_{zz}=\Sigma_{5}=\left[ 
\begin{array}{cccc}
1 & 0 & 0 & 0 \\ 
0 & -1 & 0 & 0 \\ 
0 & 0 & -1 & 0 \\ 
0 & 0 & 0 & 1%
\end{array}
\right] $ \\ 
$\Sigma_{zx}=\Sigma_{6}=\left[ 
\begin{array}{cccc}
0 & 1 & 0 & 0 \\ 
1 & 0 & 0 & 0 \\ 
0 & 0 & 0 & -1 \\ 
0 & 0 & -1 & 0%
\end{array}
\right] $ & $\Sigma_{zy}=\Sigma_{7}=\left[ 
\begin{array}{cccc}
0 & -i & 0 & 0 \\ 
i & 0 & 0 & 0 \\ 
0 & 0 & 0 & i \\ 
0 & 0 & -i & 0%
\end{array}
\right] $\  \\ 
$\Sigma_{xw}=\Sigma_{8}=\left[ 
\begin{array}{cccc}
0 & 0 & 1 & 0 \\ 
0 & 0 & 0 & 1 \\ 
1 & 0 & 0 & 0 \\ 
0 & 1 & 0 & 0%
\end{array}
\right] $ & $\Sigma_{xz}=\Sigma_{9}=\left[ 
\begin{array}{cccc}
0 & 0 & 1 & 0 \\ 
0 & 0 & 0 & -1 \\ 
1 & 0 & 0 & 0 \\ 
0 & -1 & 0 & 0%
\end{array}
\right] $ \\ 
$\Sigma_{xx}=\Sigma_{10}=\left[ 
\begin{array}{cccc}
0 & 0 & 0 & 1 \\ 
0 & 0 & 1 & 0 \\ 
0 & 1 & 0 & 0 \\ 
1 & 0 & 0 & 0%
\end{array}
\right] $ & $\Sigma_{xy}=\Sigma_{11}=\left[ 
\begin{array}{cccc}
0 & 0 & 0 & -i \\ 
0 & 0 & i & 0 \\ 
0 & -i & 0 & 0 \\ 
i & 0 & 0 & 0%
\end{array}
\right] $ \\ 
$\Sigma_{yw}=\Sigma_{12}=\left[ 
\begin{array}{cccc}
0 & 0 & -i & 0 \\ 
0 & 0 & 0 & -i \\ 
i & 0 & 0 & 0 \\ 
0 & i & 0 & 0%
\end{array}
\right] $ & $\Sigma_{yz}=\Sigma_{13}=\left[ 
\begin{array}{cccc}
0 & 0 & -i & 0 \\ 
0 & 0 & 0 & i \\ 
i & 0 & 0 & 0 \\ 
0 & -i & 0 & 0%
\end{array}
\right] $ \\ 
$\Sigma_{yx}=\Sigma_{14}=\left[ 
\begin{array}{cccc}
0 & 0 & 0 & -i \\ 
0 & 0 & -i & 0 \\ 
0 & i & 0 & 0 \\ 
i & 0 & 0 & 0%
\end{array}
\right] $ & $\Sigma_{yy}=\Sigma_{15}=\left[ 
\begin{array}{cccc}
0 & 0 & 0 & -1 \\ 
0 & 0 & 1 & 0 \\ 
0 & 1 & 0 & 0 \\ 
-1 & 0 & 0 & 0%
\end{array}
\right] $%
\end{tabular}%
\end{equation*}

The products of these matrices can easily be found from

\begin{equation*}
\Sigma_{\lambda\mu}\Sigma_{\eta\nu}=\left( \sigma_{\lambda}\otimes\sigma
_{\mu}\right) \left( \sigma_{\eta}\otimes\sigma_{\nu}\right) =\left(
\sigma_{\lambda}\sigma_{\eta}\right) \otimes\left( \sigma_{\mu}\sigma_{\nu
}\right) .
\end{equation*}
The maximal pseudostabilizers in$\ \mathcal{S}_{2}$ are presented below.%
\footnote{%
Constructing the subsets of mutually commuting Pauli matrices can be done
through a series of logical steps. \ The key is to note that $%
[\Sigma_{jk},\Sigma_{lm}]=0$ requires either $[\sigma_{j},\sigma_{l}]=[%
\sigma_{k},\sigma_{m}]=0\ $or $\{\sigma_{j},\sigma_{l}\}=\{\sigma
_{k},\sigma_{m}\}=0\ $}%
\begin{equation*}
\begin{tabular}{|c|c|c|}
\hline
Subset \# & Letter notation & Number notation \\ \hline
$1$ & $\left\{ \Sigma_{ww},\Sigma_{wz},\Sigma_{zw},\Sigma_{zz}\right\} $ & $%
\left\{ \Sigma_{0},\Sigma_{1},\Sigma_{4},\Sigma_{5}\right\} $ \\ \hline
$2$ & $\left\{ \Sigma_{ww},\Sigma_{wx},\Sigma_{zw},\Sigma_{zx}\right\} $ & $%
\left\{ \Sigma_{0},\Sigma_{2},\Sigma_{4},\Sigma_{6}\right\} $ \\ \hline
$3$ & $\{\Sigma_{ww},\Sigma_{wy},\Sigma_{zw},\Sigma_{zy}\}$ & $\left\{
\Sigma_{0},\Sigma_{3},\Sigma_{4},\Sigma_{7}\right\} $ \\ \hline
$4$ & $\left\{ \Sigma_{ww},\Sigma_{wz},\Sigma_{xw},\Sigma_{xz}\right\} $ & $%
\left\{ \Sigma_{0},\Sigma_{1},\Sigma_{8},\Sigma_{9}\right\} $ \\ \hline
$5$ & $\left\{ \Sigma_{ww},\Sigma_{wx},\Sigma_{xw},\Sigma_{xx}\right\} $ & $%
\left\{ \Sigma_{0},\Sigma_{2},\Sigma_{8},\Sigma_{10}\right\} $ \\ \hline
$6$ & $\left\{ \Sigma_{ww},\Sigma_{wy},\Sigma_{xw},\Sigma_{xy}\right\} $ & $%
\left\{ \Sigma_{0},\Sigma_{3},\Sigma_{8},\Sigma_{11}\right\} $ \\ \hline
$7$ & $\left\{ \Sigma_{ww},\Sigma_{wz},\Sigma_{yw},\Sigma_{yz}\right\} $ & $%
\left\{ \Sigma_{0},\Sigma_{1},\Sigma_{12},\Sigma_{13}\right\} $ \\ \hline
$8$ & $\left\{ \Sigma_{ww},\Sigma_{wx},\Sigma_{yw},\Sigma_{yx}\right\} $ & $%
\left\{ \Sigma_{0},\Sigma_{2},\Sigma_{12},\Sigma_{14}\right\} $ \\ \hline
$9$ & $\left\{ \Sigma_{ww},\Sigma_{wy},\Sigma_{yw},\Sigma_{yy}\right\} $ & $%
\left\{ \Sigma_{0},\Sigma_{3},\Sigma_{12},\Sigma_{15}\right\} $ \\ \hline
$10^{\ast}$ & $\left\{ \Sigma_{ww},\Sigma_{zz},\Sigma_{xx},\Sigma
_{yy}\right\} $ & $\left\{ \Sigma_{0},\Sigma_{5},\Sigma_{10},\Sigma
_{15}\right\} $ \\ \hline
$11^{\ast}$ & $\{\Sigma_{ww},\Sigma_{zz},\Sigma_{xy},\Sigma_{yx}\}$ & $%
\left\{ \Sigma_{0},\Sigma_{5},\Sigma_{11},\Sigma_{14}\right\} $ \\ \hline
$12^{\ast}$ & $\left\{ \Sigma_{ww},\Sigma_{zx},\Sigma_{xz},\Sigma
_{yy}\right\} $ & $\left\{ \Sigma_{0},\Sigma_{6},\Sigma_{9},\Sigma
_{15}\right\} $ \\ \hline
$13^{\ast}$ & $\left\{ \Sigma_{ww},\Sigma_{zx},\Sigma_{xy},\Sigma
_{yz}\right\} $ & $\left\{ \Sigma_{0},\Sigma_{6},\Sigma_{11},\Sigma
_{13}\right\} $ \\ \hline
$14^{\ast}$ & $\{\Sigma_{ww},\Sigma_{zy},\Sigma_{xz},\Sigma_{yx}\}$ & $%
\left\{ \Sigma_{0},\Sigma_{7},\Sigma_{9},\Sigma_{14}\right\} $ \\ \hline
$15^{\ast}$ & $\{\Sigma_{ww},\Sigma_{zy},\Sigma_{xx},\Sigma_{yz}\}$ & $%
\left\{ \Sigma_{0},\Sigma_{7},\Sigma_{10},\Sigma_{13}\right\} $ \\ \hline
\end{tabular}
\ \ \ 
\end{equation*}
Each of these fifteen sets, or pseudostabilizers, will yield four
simultaneous eigenvectors, contributing four states to $\mathfrak{H}_{2}$. \
We therefore recover the result that $\mathfrak{H}_{2}$ has sixty states. \ 

These subsets can be classified as corresponding to entangled or product
states by examining their generators. Note that each pseudostabilizer is
generated by any two of its non-identity elements. The presence of the
one-qubit identity $\sigma_{w}$ when the generators are decomposed into
tensor products of one-qubit Pauli matrices implies that the states
corresponding to that subset are product states. Conversely, the absence of
the identity in this decomposition indicates that the states corresponding
to that Abelian subset are fully entangled states. The subsets whose
generators do not contain the one-qubit identity are denoted here and below
by an asterisk($\ast$).\FRAME{ftbpFU}{4.9139in}{4.1706in}{0pt}{\Qcb{Graph of
the set $\mathcal{S}_{2}$ of two-qubit generalized Pauli matrices $%
\Sigma_{j} $ (circles bearing the subscript of the matrix in letter
notation) and by the pseudostabilizers (triangles formed by three connected
circles). The three $\Sigma^{\prime}s$ in a triangle share four common
eigenvectors which form an orthonormal basis spanning the two-qubit Hilbert
space. The fifteen triangles thus give fifteen sets of four basis vectors.
Shaded triangles correspond to entangled states while non-shaded triangles
correspond to product states. Neighbouring triangles have one (non-identity) 
$\Sigma$ in common, and each (non-identity) $\Sigma$ is shared by three
triangles. The line segments joining the vertices of a triangle correspond
to pairs $\left\{ j,k\right\} $ of commuting matrices; each segment
therefore specifies a $\protect\pi/2$ rotation $X_{j,k}=(\Sigma_{j}+i%
\Sigma_{k})/\protect\sqrt{2}$ that transforms the eigenvectors of an
adjacent triangle into its neighbour. The figure thus constitutes a
\textquotedblleft roadmap\textquotedblright\ for navigating the discrete set 
$\mathfrak{H}_{2}$. Repeated circles indicate the closure of the graph.}}{}{%
pseudostabilizerseps.eps}{\special{ language "Scientific Word"; type
"GRAPHIC"; maintain-aspect-ratio TRUE; display "USEDEF"; valid_file "F";
width 4.9139in; height 4.1706in; depth 0pt; original-width 4.8594in;
original-height 4.1208in; cropleft "0"; croptop "1"; cropright "1";
cropbottom "0"; filename '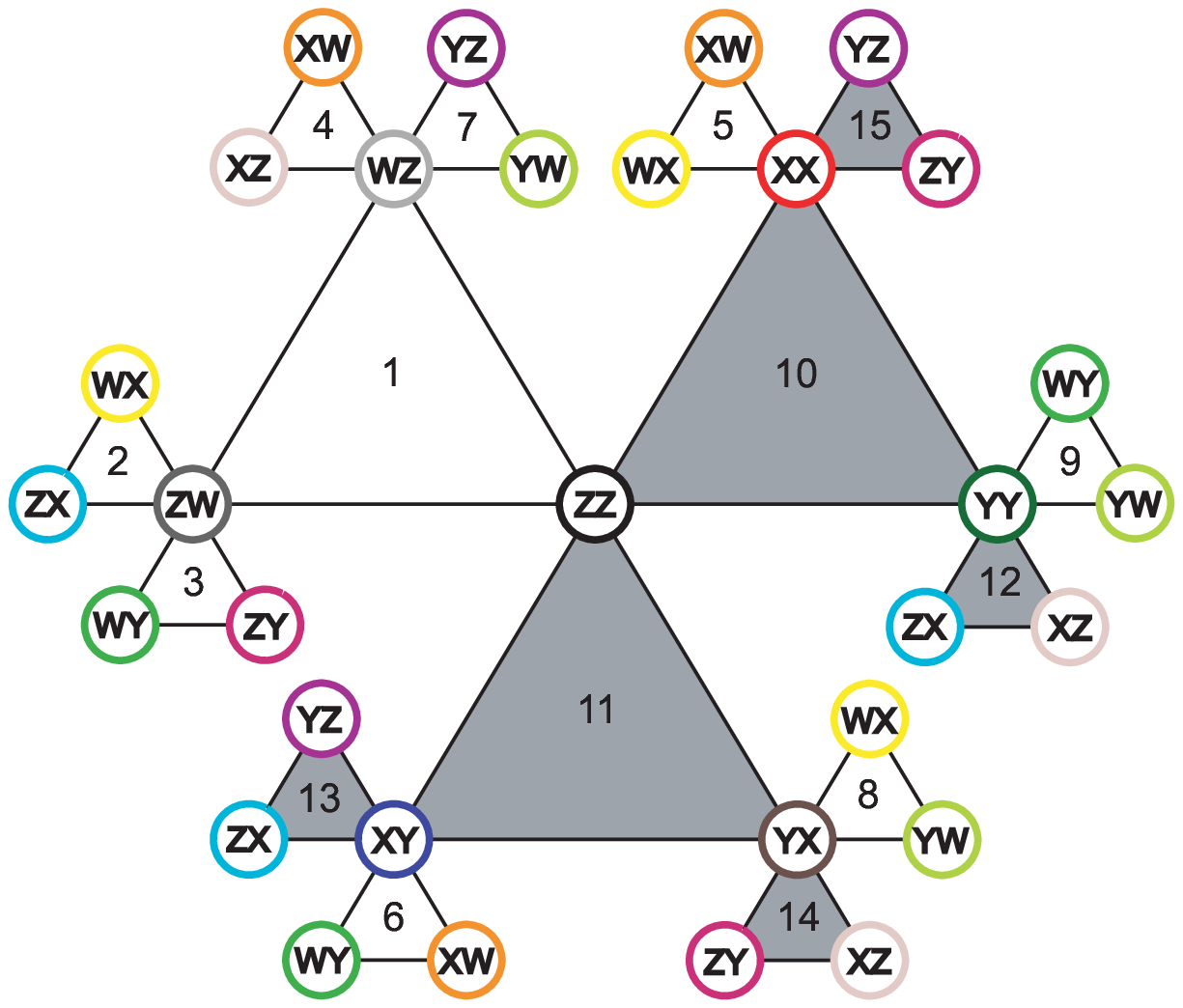';file-properties "XNPEU";}%
}

We can obtain all the states of $\mathfrak{H}_{2}$ directly by forming a
mixed linear combination of the first two non-identity elements from within
each set. For instance $X_{3,12}=\frac{1}{\sqrt{2}}\left(
\Sigma_{wy}+i\Sigma _{yw}\right) $, when diagonalized, gives four orthogonal
eigenvectors with four different eigenvalues. (Note that other rotations
from $s_{2}^{1}$, such as $X_{12\,,15}=\frac{1}{\sqrt{2}}\left(
\Sigma_{12}+i\Sigma_{15}\right) $ and $X_{3,15}=\frac{1}{\sqrt{2}}\left(
\Sigma_{3}+i\Sigma_{15}\right) $ will produce the same four eigenvectors,
but with permuted eigenvalues.) We thus construct in this manner fifteen
generalized $X^{\prime}s,$ each having a different principal axis, which
form the seed of $\mathfrak{H}_{2}.$ Diagonalization of the seed $%
X^{\prime}s $ exhaustively gives the eigenvectors constituting $\mathfrak{H}%
_{2}$. We list these eigenvectors below, separated into columns
corresponding to the eigenvalues $(\pm1\pm i)$. For clarity, we list them as
unnormalized row vectors. Again, entangled states are denoted by an asterisk
on the set label.%
\begin{equation*}
\begin{tabular}{|c|c|c|c|c|}
\hline
Set & $-1-i$ & $-1+i$ & $1-i$ & $1+i$ \\ \hline
$1$ & $\left( 0,0,0,1\right) $ & $\left( 0,1,0,0\right) $ & $\left(
0,0,1,0\right) $ & $\left( 1,0,0,0\right) $ \\ \hline
$2$ & $\left( 0,0,-1,1\right) $ & $\left( -1,1,0,0\right) $ & $\left(
0,0,1,1\right) $ & $\left( 1,1,0,0\right) $ \\ \hline
$3$ & $\left( 0,0,i,1\right) $ & $\left( i,1,0,0\right) $ & $\left(
0,0,-i,1\right) $ & $\left( -i,1,0,0\right) $ \\ \hline
$4$ & $\left( 0,-1,0,1\right) $ & $\left( 0,1,0,1\right) $ & $\left(
-1,0,1,0\right) $ & $\left( 1,0,1,0\right) $ \\ \hline
$5$ & $\left( 1,-1,-1,1\right) $ & $\left( -1,1,-1,1\right) $ & $\left(
-1,-1,1,1\right) $ & $\left( 1,1,1,1\right) $ \\ \hline
$6$ & $\left( -i,-1,i,1\right) $ & $\left( i,1,i,1\right) $ & $\left(
i,-1,-i,1\right) $ & $\left( -i,1,-i,1\right) $ \\ \hline
$7$ & $\left( 0,i,0,1\right) $ & $\left( 0,-i,0,1\right) $ & $\left(
i,0,1,0\right) $ & $\left( -i,0,1,0\right) $ \\ \hline
$8$ & $\left( -i,i,-1,1\right) $ & $\left( i,-i,-1,1\right) $ & $\left(
i,i,1,1\right) $ & $\left( -i,-i,1,1\right) $ \\ \hline
$9$ & $\left( -1,i,i,1\right) $ & $\left( 1,-i,i,1\right) $ & $\left(
1,i,-i,1\right) $ & $\left( 1,i,i,-1\right) $ \\ \hline
$10^{\ast}$ & $\left( 0,-1,1,0\right) $ & $\left( -1,0,0,1\right) $ & $%
\left( 1,0,0,1\right) $ & $\left( 0,1,1,0\right) $ \\ \hline
$11^{\ast}$ & $\left( i,0,0,1\right) $ & $\left( 0,-i,1,0\right) $ & $\left(
0,i,1,0\right) $ & $\left( -i,0,0,1\right) $ \\ \hline
$12^{\ast}$ & $\left( 1,1,-1,1\right) $ & $\left( -1,1,1,1\right) $ & $%
\left( 1,-1,1,1\right) $ & $\left( 1,1,1,-1\right) $ \\ \hline
$13^{\ast}$ & $\left( i,-i,1,1\right) $ & $\left( i,i,-1,1\right) $ & $%
\left( i,i,1,-1\right) $ & $\left( -i,i,1,1\right) $ \\ \hline
$14^{\ast}$ & $\left( i,1,-i,1\right) $ & $\left( -i,1,i,1\right) $ & $%
\left( i,-1,i,1\right) $ & $\left( i,1,i,-1\right) $ \\ \hline
$15^{\ast}$ & $\left( -1,-i,i,1\right) $ & $\left( -1,i,-i,1\right) $ & $%
\left( 1,-i,-i,1\right) $ & $\left( 1,i,i,1\right) $ \\ \hline
\end{tabular}
\ \ 
\end{equation*}

Note that the first stabilizer of the product sector corresponds to the
computational basis, while the first stabilizer of the entangled sector
corresponds to the Bell basis.

This method not only finds the states of $\mathfrak{H}_{2}$ in an exhaustive
way. It also provides a road map for\ navigating the discrete set. To
illustrate this, consider three pseudostabilizers which we call $s^{a}$, $%
s^{b}$ and $s^{c}$. They have two generalized Pauli matrices in common, one
of them being the trivial $\Sigma_{0}$. We chose one of the two and call it $%
\Sigma_{m}$. Consider three $\Sigma\neq\Sigma_{m},$%
\begin{equation*}
\begin{tabular}{lll}
$\Sigma_{j}\in s^{a},$ & $\Sigma_{k}\in s^{b},$ & $and$ $\ \ \ \Sigma_{l}\in
s^{c}$%
\end{tabular}
\ \ \ 
\end{equation*}
such that

\begin{equation*}
\left\{ \Sigma_{j},\Sigma_{k}\right\} =\left\{ \Sigma_{k},\Sigma
_{l}\right\} =\left\{ \Sigma_{l},\Sigma_{j}\right\} =0.
\end{equation*}
Since all pairs of generalized Pauli matrices that do not commute must
anti-commute, they also satisfy

\begin{equation*}
\left[ \Sigma_{j},\Sigma_{m}\right] =\left[ \Sigma_{k},\Sigma_{m}\right] =%
\left[ \Sigma_{l},\Sigma_{m}\right] =0.
\end{equation*}
Then it is easy to show by a direct calculation that

\begin{align*}
X_{m\,j}^{a} & =\frac{1}{\sqrt{2}}\left( \Sigma_{m}+i\Sigma_{j}\right) , \\
X_{m\,k}^{b} & =\frac{1}{\sqrt{2}}\left( \Sigma_{m}+i\Sigma_{k}\right) , \\
X_{m\,l}^{c} & =\frac{1}{\sqrt{2}}\left( \Sigma_{m}+i\Sigma_{l}\right) ,
\end{align*}
have one of the properties:

\begin{equation*}
X_{m\,k}^{b}X_{m\,j}^{a}X_{k\,m}^{b}=X_{m\,l}^{c}\text{ }
\end{equation*}
or 
\begin{equation*}
X_{m\,k}^{b}X_{m\,j}^{a}X_{k\,m}^{b}=(X_{m\,l}^{c})^{-1}=-iX_{l\,m}^{c}\text{
}
\end{equation*}
depending on whether the two anti-commuting $\sigma^{\prime}s$ in the
decomposition of $\Sigma_{j}\Sigma_{k}$ appear in cyclic order or
anti-cyclic order, respectively. This means, following the proof given for
the one-qubit case, that all the eigenvectors of $s^{a}$ are transformed
into the eigenvectors of $s^{c}$ by the transformation $X^{b}$.

All together, there are $120$ different generalized two-qubit $\pi/2$
rotations generated by the scheme 
\begin{equation*}
X_{i\text{ }j}=\frac{1}{\sqrt{2}}\left( \Sigma_{i}+i\Sigma_{j}\right) ,
\end{equation*}
where $\Sigma_{i}$ and $\Sigma_{j}$ commute.

Among these $X^{\prime}s$, there is a subset that plays an important
practical role. These are $\pi/2$ rotations of the form

\begin{equation*}
X_{0\text{ }j}=\frac{1}{\sqrt{2}}\left( \Sigma_{0}+i\Sigma_{j}\right) \text{.%
}
\end{equation*}
They correspond to the unitary time-evolution operator

\begin{equation*}
U\left( t\right) =e^{i\Sigma_{j}\tau}
\end{equation*}
with $\tau=\pi/4$, and are thus directly implemented by a Hamiltonian
proportional to $\Sigma_{j}$. These rotations constitute the practical means
of navigating $\mathfrak{H}_{2}$. They can be seen as the \textquotedblleft
primitives\textquotedblright\ of the Clifford group, as we show below.

But first, it is important to note that the generalized Pauli matrices in
the above arguments are not limited to the two-qubit case, but can in fact
be over any number of qubits. These results are therefore directly
generalizable to larger Hilbert spaces $H_{N}$ and larger discrete sets $%
\mathfrak{H}_{N}$. We duly conclude that our generalized $\pi/2$ rotations
on $N$ qubits, constructed from the pseudostabilizers $s_{N}^{a}$, leave $%
\mathfrak{H}_{N}$ invariant.

\subsection{The generalized $\protect\pi/2$ rotations generate the $N$-qubit
Clifford group}

So far we have successfully discretized the continuous Hilbert space $H_{N}$%
, and in doing so we have described a class of generalized $\pi/2$ rotations
that leave the $\mathfrak{H}_{N}$ invariant. From the point of view of
operators acting in $H_{N}$, this discretization means we have reduced the
continuous transformation group $SU(2^{N})$ to a finite group. Here we show
that this finite group is the $N$-qubit Clifford group $\mathcal{C}_{N}.$

The Clifford group is defined as the normalizer of the Pauli group. That is,
a unitary operator $X$ is contained in $\mathcal{C}_{N}$ if and only if 
\begin{equation*}
X\Sigma X^{-1}\in\mathcal{G}_{N}\;\forall\text{ }\Sigma\in\mathcal{G}_{N}.
\end{equation*}

First, let us show that our $X^{\prime}s$ are elements of $\mathcal{C}_{N}$.
That is, 
\begin{equation*}
X_{j\text{ }k}=\frac{1}{\sqrt{2}}(\Sigma_{j}+i\Sigma_{k})\in\mathcal{C}%
_{N}\;\ \text{if }\left[ \Sigma_{j},\Sigma_{k}\right] =0.
\end{equation*}
We have

\begin{align*}
\Sigma_{j}\Sigma_{l} & =\varepsilon_{jl}\Sigma_{l}\Sigma_{j}, \\
\Sigma_{k}\Sigma_{l} & =\varepsilon_{kl}\Sigma_{l}\Sigma_{k},
\end{align*}
where $\varepsilon_{jl}=\pm1$ and $\varepsilon_{kl}=\pm1.$Thus,

\begin{align*}
X_{j\text{\thinspace}k}\Sigma_{l}X_{j\text{\thinspace}k}^{-1} & =\frac{1}{2}%
\left( \Sigma_{j}+i\Sigma_{k}\right) \Sigma_{l}\left(
\Sigma_{j}-i\Sigma_{k}\right) \\
& =\frac{1}{2}\Sigma_{l}\left( \varepsilon_{jl}\Sigma_{j}+i\varepsilon
_{kl}\Sigma_{k}\right) \left( \Sigma_{j}-i\Sigma_{k}\right) \\
& =\frac{1}{2}\varepsilon_{jl}\Sigma_{l}\left( \Sigma_{j}+i\varepsilon
_{kl}\varepsilon_{jl}\Sigma_{k}\right) \left( \Sigma_{j}-i\Sigma_{k}\right) .
\end{align*}
If $\varepsilon_{kl}\varepsilon_{jl}=1,$ 
\begin{align*}
& =\frac{1}{2}\varepsilon_{jl}\Sigma_{l}\left(
2\Sigma_{0}+i\Sigma_{k}\Sigma_{j}-i\Sigma_{j}\Sigma_{k}\right) \\
& =\varepsilon_{jl}\Sigma_{l}\in\mathcal{G}_{N}.
\end{align*}
If $\varepsilon_{kl}\varepsilon_{jl}=-1,$%
\begin{align*}
& =\frac{1}{2}\varepsilon_{jl}\Sigma_{l}\left( -i\Sigma_{k}\Sigma
_{j}-i\Sigma_{j}\Sigma_{k}\right) \\
& =-i\varepsilon_{jl}\Sigma_{l}\Sigma_{j}\Sigma_{k}\in\mathcal{G}_{N}.
\end{align*}
So the generalized $\pi/2$ rotations on $N$ qubits are elements of the
Clifford group.

Now note that the Clifford group is generated by the Hadamard, 
\begin{equation*}
\begin{tabular}{l}
$\QTR{sc}{H}=\frac{1}{\sqrt{2}}%
\begin{bmatrix}
1 & 1 \\ 
1 & -1%
\end{bmatrix}
$%
\end{tabular}
\ \ \ 
\end{equation*}
phase, 
\begin{equation*}
\QTR{sc}{S}=%
\begin{bmatrix}
1+i & 0 \\ 
0 & 1-i%
\end{bmatrix}%
\end{equation*}
and CNOT,%
\begin{equation*}
\QTR{sc}{CNOT}=%
\begin{bmatrix}
1 & 0 & 0 & 0 \\ 
0 & 1 & 0 & 0 \\ 
0 & 0 & 0 & 1 \\ 
0 & 0 & 1 & 0%
\end{bmatrix}%
\end{equation*}
gates\cite{Nielsen-Chuang}. The Hadamard gate may be composed from the
one-qubit $\pi/2$ rotations $X_{0\,2}$ and $X_{0\,1}$:%
\begin{align*}
\QTR{sc}{H} & =X_{0\,2}X_{0\,1}(X_{2\text{\thinspace}0})^{-1} \\
& =-iX_{0\,2}X_{0\,1}X_{0\text{\thinspace}2} \\
& =\frac{-i}{2\sqrt{2}}%
\begin{bmatrix}
1 & i \\ 
i & 1%
\end{bmatrix}%
\begin{bmatrix}
1+i & 0 \\ 
0 & 1-i%
\end{bmatrix}%
\begin{bmatrix}
1 & i \\ 
i & 1%
\end{bmatrix}
\\
& =\frac{1}{\sqrt{2}}%
\begin{bmatrix}
1 & 1 \\ 
1 & -1%
\end{bmatrix}
.
\end{align*}
The phase gate may be trivially constructed from a single one-qubit $\pi/2$
rotation:%
\begin{align*}
\QTR{sc}{S} & =X_{0\,1} \\
& =%
\begin{bmatrix}
1+i & 0 \\ 
0 & 1-i%
\end{bmatrix}
,
\end{align*}
while the CNOT\ is simply the product of three $X^{\prime}s:$%
\begin{align*}
\QTR{sc}{CNOT} & =(X_{0\,2})^{-1}X_{0\,\,6}(X_{0\,\,4})^{-1} \\
& =-X_{2\,0}X_{0\,\,6}X_{4\,0} \\
& =\frac{-1}{2\sqrt{2}}%
\begin{bmatrix}
i & 1 & 0 & 0 \\ 
1 & i & 0 & 0 \\ 
0 & 0 & i & 1 \\ 
0 & 0 & 1 & i%
\end{bmatrix}
\left[ 
\begin{array}{cccc}
1 & i & 0 & 0 \\ 
i & 1 & 0 & 0 \\ 
0 & 0 & 1 & -i \\ 
0 & 0 & -i & 1%
\end{array}
\right] 
\begin{bmatrix}
1+i & 0 & 0 & 0 \\ 
0 & 1+i & 0 & 0 \\ 
0 & 0 & -1+i & 0 \\ 
0 & 0 & 0 & -1+i%
\end{bmatrix}
\\
& =\frac{1}{\sqrt{2}}%
\begin{bmatrix}
1-i & 0 & 0 & 0 \\ 
0 & 1-i & 0 & 0 \\ 
0 & 0 & 0 & 1-i \\ 
0 & 0 & 1-i & 0%
\end{bmatrix}
.
\end{align*}
So our generalized $\pi/2$ rotations allow a direct construction of a gate
set that generates the Clifford group. The finite transformation group
leaving $\mathfrak{H}_{N}$ invariant, generated by the generalized $\pi/2$
rotations on $N$ qubits, is thus the $N$-qubit Clifford group $\mathcal{C}%
_{N}$.

\subsection{Comments}

One of the motives we presented for this work was the difficulty we
anticipate in achieving the reliability requisite for fault-tolerant quantum
computation. Clearly, limiting the register to a finite number of possible
states must alleviate this difficulty, but by how much?

It can be derived from the properties of the Pauli group that each
pseudostabilizer $s_{N}^{a}$ has $N$\ levels of non-orthogonal neighbors.
Since the eigenstates of neighboring pseudostabilizers are connected by a
single $\pi/2$ rotation, any state on $\mathfrak{H}_{N}$ can be reached from
any other in at most $N+1$ such rotations. This is to be compared with the
result that an arbitrary state in the full Hilbert space can be reached to
within an error $\epsilon$ by concatenating $\mathcal{O}(\log^{c}(1/%
\epsilon))$ rotations from a universal set, with $c\approx2$. In addition,
note that the \textsc{CNOT} and \textsc{H} gates are not directly
implemented by a physical Hamiltonian, but must be built up from $\pi/2$
rotations which are naturally realized with accessible field variations, so
there is a second simplification from working with the $\pi/2$ rotations
rather than standard universal gate sets such as \textsf{\{}\textsc{H}%
\textsf{, }\textsc{S}\textsf{, }\textsc{T}\textsf{, }\textsc{CNOT}\textsf{\}.%
}

From an experimentalist's point of view, therefore, the $X^{\prime}s$ form a
very natural language for building quantum gates. A rotation of the form $%
X_{0\text{\thinspace}j}=\left( \Sigma_{0}+i\Sigma_{j}\right) /\sqrt{2}$ is
directly implemented by a term in the Hamiltonian proportional to $%
\Sigma_{j} $. And as shown above, this class of rotations generates $%
\mathcal{C}_{N}$. The $X^{\prime}s$ are thus the basic instructions for a
sort of \textquotedblleft machine language\textquotedblright\ for quantum
processors. The following section shows a simple example of their calculus.

\subsection{Sample application of digital quantum information: implementing 
\textsc{CNOT}}

The Hamiltonian describing a given physical system determines which of the
generalized $\pi/2$ rotations will be directly realizable in that system.
Implementing \textsc{CNOT} according to the decomposition in section 2.5
requires a physical system with a Hamiltonian proportional to $\Sigma_{zx}$
in order to realize the entangling operation $X_{0\,\,6}=(\Sigma_{ww}+i%
\Sigma_{zx})/\sqrt{2}$. Though this type of inter-qubit interaction is
possible\footnote{%
As a charge-flux coupling between superconducting qubits, for example.},
most qubit systems rely on a less exotic interaction, such as one
proportional to $\Sigma_{zz}$ or $\Sigma_{xx}$. How can we implement the 
\textsc{CNOT} gate in one of these more standard registers?

Specifically, suppose the system is described by a two-qubit Hamiltonian of
the form 
\begin{equation*}
H=a(t)\Sigma_{1}+b(t)\Sigma_{2}+c(t)\Sigma_{4}+d(t)\Sigma_{8}+e(t)\Sigma_{5},
\end{equation*}
where the tuning parameters $a,b,c,d,e$ allow the relative strengths of the
terms to be adjusted during an experiment. Our task is to replace the
rotation $X_{0\,\,6}$ in the sequence $(X_{0\,2})^{-1}X_{0\,\,6}(X_{0\,%
\,4})^{-1}$ with a rotation or sequence of rotations generated by the above
Hamiltonian. Following the discussion in sections 2.3 and 2.4 on the
relationships between the generalized $\pi/2$ rotations, it is
straightforward to calculate that $X_{0\,\,6}=(X_{0\,\,3})^{-1}X_{0\,%
\,5}X_{0\,\,3}$ while $X_{0\,\,3}=(X_{0\,\,1})^{-1}X_{0\,\,2}X_{0\,\,1}$.
Together these give an alternate decomposition that employs only directly
realizable $X^{\prime}s$ :%
\begin{align*}
\QTR{sc}{CNOT} & =(X_{0\,2})^{-1}X_{0\,\,6}(X_{0\,\,4})^{-1} \\
& =(X_{0\,2})^{-1}(X_{0\,\,3})^{-1}X_{0\,\,5}X_{0\,\,3}(X_{0\,\,4})^{-1} \\
&
=(X_{0\,\,2})^{-1}X_{0\,\,1}X_{0\,\,2}(X_{0\,\,1})^{-1}X_{0\,\,5}(X_{0\,%
\,1})^{-1}X_{0\,\,2}X_{0\,\,1}(X_{0\,\,4})^{-1}.
\end{align*}
\qquad Since the decompositions differ in their implementation but not in
their meaning, such sequences are \textquotedblleft
synonyms.\textquotedblright\ It is important to note that the \textsc{CNOT}
is not a special case: a synonym suitable for a particular implementation
could likewise be calculated for \textit{any} gate in $C_{2}.$

\subsection{Conclusion to section 2}

In this section we have presented a geometric method for producing from the
continuous Hilbert space $H_{N}$ a discrete, uniform sampling $\mathfrak{H}%
_{N}$. Because all the states in the discrete set are geometrically
equivalent, $\mathfrak{H}_{N}$ represents a generalized Platonic solid in $%
H_{N}.$ This method is closely related to the stabilizer formalism of
quantum error-correcting codes. Inherent in our construction is a
description of how different elements of $\mathfrak{H}_{N}$ are related by
transformations generated by physical Hamiltonians expressed in the basis of
generalized Pauli matrices. This has been demonstrated in detail for $%
\mathfrak{H}_{1}$ and $\mathfrak{H}_{2}$, and is obtainable by direct
analogy for higher-dimensional spaces. These ideas provide a useful tool for
analyzing problems in traditional quantum computation, as the example above
illustrates. And though computation over the discrete set $\mathfrak{H}_{N}$
is clearly less powerful than computation in the full $H_{N}$, it is
potentially more powerful than classical computation.

\section{An alternate approach to discretization: shelling the
high-dimensional dense lattices}

In this last section, we present an alternate approach to discretization
that addresses the Hilbert space directly, without reference to operators in 
$H_{N}$ or the relevant transformation groups.

Our strategy here is the following (we note $n=2^{N}$): the normalization
condition, together with the writing of complex numbers as pairs of real
numbers, identifies the Hilbert space $H_{N}$ to the high-dimensional sphere 
$S^{2n-1}$ embedded in $%
\mathbb{R}
^{2n}$. In order to discretize these hyperspheres, we use the successive
shells of dense lattices in $%
\mathbb{R}
^{2n}$. At the same time, we must take into account the global phase
freedom, and show how a discretization of the projective Hilbert space is
induced. This means that several points on $S^{2n-1}$ will represent the
same physical state, as explained below. \ In light of this, it is important
to distinguish between \textquotedblleft qubit
states\textquotedblright---the quantum states associated to the points on $%
S^{2n-1}$---and \textquotedblleft physical states\textquotedblright---the
states in the projective Hilbert space, which has the geometry of a complex
projective space $CP^{n-1}$.

As in section 2, we again focus on the one- and two-qubit cases. The
discretization of $H_{1}$is first presented in terms of the $24$ vertices of
a self-dual polytope on $S^{3},$ denoted $\left\{ 3,4,3\right\} $, which is
the first shell of the densest packing in $%
\mathbb{R}
^{4}$, denoted $\Lambda_{4}$. These $24$ vertices corresponds to $24$
one-qubit states, and, modulo a global phase, to $6$ physical states. For
the two-qubit case, we use the \textit{Gosset polytope}, the first shell of
the densest packing \ in $%
\mathbb{R}
^{8}$, denoted $E_{8}$. We find that this polytope has $240$ vertices
corresponding to $240$ two-qubit states, which leads to $60$ physical
states: $\ 36$ separable states and $24$ maximally entangled states, just as
in section 2.

\bigskip

The $N$-qubit Hilbert space is high-dimensional, and its multipartite nature
(it is the tensor product of single-qubit Hilbert space) induces a subtle
structure related to the state's various levels of entanglement\cite%
{mosseri-dandolof,karol}, which is not fully understood for $N\geq3$. The
Hopf map that is used here in the two-qubit case is entanglement sensitive,
which translates here in grouping sets of equally entangled qubit states.

\subsection{The one-qubit case}

The generic one-qubit state reads 
\begin{equation*}
\left\vert \Psi\right\rangle =t_{0}\left\vert 0\right\rangle
+t_{1}\left\vert 1\right\rangle ,
\end{equation*}
with $\left\vert t_{0}\right\vert ^{2}+\left\vert t_{1}\right\vert ^{2}=1.$%
The normalization condition identifies the set of normalized states to a
sphere $S^{3}$ embedded in $%
\mathbb{R}
^{4}.$ The projective case---the set of states modulo a global phase---leads
to the Bloch sphere description, which can be seen as the base of the $S^{3}$
Hopf fibration,\cite{urtbanke,sadoc-mosseri-livre}. An interesting discrete
model on $S^{3}$ is provided by the self-dual $\left\{ 3,4,3\right\} $
polytope \cite{coxeter-73}. It is related to the \textquotedblleft
Hurwitz\textquotedblright\ quaternion group. We now give two possible (dual)
coordinates for its vertices, in each case as a real quadruplet and a
complex pair. The correspondence between real quadruplets and complex pairs
amounts simply to taking the first two (last two) real numbers as the real
and imaginary part of the first (second) complex number. The first (second)
complex number in the pair corresponds to $t_{0}$ ($t_{1}$).

A first set, denoted $T_{1},$ is the union of the eight permutations of type 
$(\pm1,0,0,0)$ and the sixteen permutations of type $%
{\frac12}%
(\pm1,\pm1,\pm1,\pm1)$. Note that, modulo a global phase factor, these
twenty-four points really represent six different physical states, which
appear on the Bloch sphere as opposite points on the three orthogonal axes $%
x,y,z$. Indeed, the four points,

\begin{equation*}
\begin{tabular}{|c|c|}
\hline
$\text{Real quadruplets}$ & $\text{Complex pairs}$ \\ \hline
$\left( 1,0,0,0\right) $ & $\left( 1,0\right) $ \\ \hline
$\left( -1,0,0,0\right) $ & $\left( -1,0\right) $ \\ \hline
$\left( 0,1,0,0\right) $ & $\left( i,0\right) $ \\ \hline
$\left( 0,-1,0,0\right) $ & $\left( -i,0\right) $ \\ \hline
\end{tabular}%
\end{equation*}
represent the states $\left\vert \Psi_{1},\omega\right\rangle =e^{i\omega
}\left\vert 0\right\rangle $, with $\omega=0,\pi/2,\pi,3\pi/2$ ,which map to
the same point on the Bloch sphere (the north pole), and they are therefore
associated to the physical state $\left\vert \Psi_{1}\right\rangle .$
Equivalently, the four points 
\begin{equation*}
\begin{tabular}{|c|c|}
\hline
$\text{Real quadruplets}$ & $\text{Complex pairs}$ \\ \hline
$\left( 0,0,1,0\right) $ & $\left( 0,1\right) $ \\ \hline
$\left( 0,0,-1,0\right) $ & $\left( 0,-1\right) $ \\ \hline
$\left( 0,0,0,1\right) $ & $\left( 0,i\right) $ \\ \hline
$\left( 0,0,0,-1\right) $ & $\left( 0,-i\right) $ \\ \hline
\end{tabular}%
\end{equation*}
represent the four states $\left\vert \Psi_{2},\omega\right\rangle
=e^{i\omega}\left\vert 1\right\rangle $ with $\omega=0,\pi/2,\pi,3\pi/2$.
The other sixteen vertices represent four other physical states, in the
following way:%
\begin{equation*}
\begin{tabular}{ll}
$\left\vert \Psi_{3}\right\rangle \equiv\frac{e^{i\left( \omega+\pi/4\right)
}}{\sqrt{2}}\left( \left\vert 0\right\rangle -\left\vert 1\right\rangle
\right) ,\;$ & $\left\vert \Psi_{4}\right\rangle \equiv\frac{e^{i\left(
\omega+\pi/4\right) }}{\sqrt{2}}\left( \left\vert 0\right\rangle -\left\vert
1\right\rangle \right) ,\;$ \\ 
$\left\vert \Psi_{5}\right\rangle \equiv\frac{e^{i\left( \omega+\pi/4\right)
}}{\sqrt{2}}\left( \left\vert 0\right\rangle +i\left\vert 1\right\rangle
\right) ,$ & $\left\vert \Psi_{6}\right\rangle \equiv\frac{e^{i\left(
\omega+\pi/4\right) }}{\sqrt{2}}\left( \left\vert 0\right\rangle
-i\left\vert 1\right\rangle \right) ,$%
\end{tabular}%
\end{equation*}
with $\omega=0,\pi/2,\pi,3\pi/2.$

For the later purpose of a discrete two-qubit construction,\ it is useful to
describe a second version of the polytope $\{3,4,3\}$, for which the
twenty-four vertices form a set $T_{2}$ given by twenty-four permutations of
the type $\left\{ \pm1,\pm1,0,0\right\} /\sqrt{2}.$ This polytope is
obtained from the former one through a \textit{screw }motion on $S^{3}$of
angle $\pi/4.$ This set leads to twenty-four states 
\begin{equation*}
\left\vert \Phi_{l},\omega\right\rangle =\epsilon\left\vert
\Psi_{l},\omega\right\rangle ,\;\;l=1..6,\;\;\omega=0,\pi/2,\pi,3\pi/2\text{%
, and \ }\epsilon=e^{i\pi/4}
\end{equation*}
and to the six one-qubit physical states $\left\vert \Phi_{l}\right\rangle $
identical to $\left\vert \Psi_{l}\right\rangle .$ Indeed, the six states $%
\left\vert \Psi_{j}\right\rangle $ sit at the vertices of a regular
octahedron. Since the states $\left\vert \Phi_{l},\omega\right\rangle $ only
differ from $\left\vert \Psi_{l},\omega\right\rangle $ by a global phase,
they map onto the same six points on the Bloch sphere.

\subsection{The two-qubit case}

We now consider the two-qubit case, for which a generic state reads: 
\begin{equation*}
\left\vert \Psi\right\rangle =t_{00}\left\vert 00\right\rangle
+t_{01}\left\vert 01\right\rangle +t_{10}\left\vert 10\right\rangle
+t_{11}\left\vert 11\right\rangle ,\;\ \text{and \ }\left\vert
t_{00}\right\vert ^{2}+\left\vert t_{01}\right\vert ^{2}+\left\vert
t_{10}\right\vert ^{2}+\left\vert t_{11}\right\vert ^{2}=1.
\end{equation*}
The normalization condition identifies the set of normalized states to the
sphere $S^{7}$ embedded in $%
\mathbb{R}
^{8}$.

As for the one-qubit case, we consider the first shell of points in the
densest lattice in $%
\mathbb{R}
^{8}$, denoted $E_{8}$. This lattice belongs to the family of laminated
lattices $\Lambda_{i}$, and is therefore sometimes denoted $\Lambda_{8}$.
These laminated lattices form a series which starts with the triangular
lattice in $2d$ (the densest lattice in $2d$). $\Lambda_{3}$ is obtained as
a particular sequence of $\Lambda_{2}$ lattices packed in a third dimension,
which gives the face centered cubic lattice, one of the two densest lattices
in $3d$. Appropriately packing $\Lambda_{3}$ lattices along a fourth
dimension leads to $\Lambda_{4}$, whose first shell is precisely the $%
\{3,4,3\}$ polytope we used above. Upon iteration, this construction
eventually leads to the $\Lambda_{8}=E_{8}$ lattice suitable for the
two-qubit case. We shall focus here on the set of $240$ sites belonging to
the $E_{8}$ first shell that forms the Gosset polytope and, as for the
one-qubit case, enumerate the physical states they represent.

\subsubsection{Discrete Hopf fibration for the Gosset polytope on $S^{7}$}

The 240 vertices of the Gosset polytope belong to the sphere $S^{7}$. These
240 vertices\ may be separated into ten equivalent subsets, each belonging
to non-intersecting $S^{3}$ spheres. This is nothing but a discrete version
of the $S^{7}$ Hopf fibration, with fibers $S^{3}$ and base $S^{4}$\cite%
{mosseri-dandolof,sadoc-mosseri-livre,sadoc-mosseri-E8,manton-87}.

It is simpler to use here quaternionic coordinates instead of complex or
real ones. The above set $T_{1}$, scaled such that the corresponding points
belong to a sphere $S^{3}$ of radius $\frac{1}{\sqrt{2}}$, now reads: 
\begin{equation*}
T_{1}=\{\pm\frac{1}{\sqrt{2}},\pm\frac{\mathbf{i}}{\sqrt{2}},\pm \frac{%
\mathbf{j}}{\sqrt{2}},\pm\frac{\mathbf{k}}{\sqrt{2}},\frac{1}{2\sqrt{2}}%
(\pm1\pm\mathbf{i}\pm\mathbf{j}\pm\mathbf{k})\},
\end{equation*}
where $\mathbf{i},\mathbf{j}$ and $\mathbf{k}$ are the standard unit
quaternions. The set $T_{2}$ stays on a unit sphere and reads: 
\begin{equation*}
T_{2}=\{\frac{1}{\sqrt{2}}(\pm1\pm\mathbf{i}),\frac{1}{\sqrt{2}}(\pm 1\pm%
\mathbf{j}),\frac{1}{\sqrt{2}}(\pm1\pm\mathbf{k}),\frac{1}{\sqrt{2}}(\pm%
\mathbf{i}\pm\mathbf{j}),\frac{1}{\sqrt{2}}(\pm\mathbf{i}\pm \mathbf{k}),%
\frac{1}{\sqrt{2}}(\pm\mathbf{j}\pm\mathbf{k})\}.
\end{equation*}
The 240 vertices of the Gosset polytope belong to the ten sets: 
\begin{gather*}
S_{1}=(T_{2},0),\text{ \ \ }S_{2}=(0,T_{2}),\text{ \ \ }S_{3}=(T_{1},T_{1}),%
\text{ \ \ }S_{4}=(T_{1},-T_{1}),\ \ \newline
\ S_{5}=(T_{1},\mathbf{i}T_{1}),\text{ \ } \\
S_{6}=(T_{1},-\mathbf{i}T_{1}),\text{ \ \ }S_{7}=(T_{1},\mathbf{j}T_{1}),%
\text{ \ \ }S_{8}=(T_{1},-\mathbf{j}T_{1}),\text{ \ \ }S_{9}=(T_{1},\mathbf{k%
}T_{1}),\text{ \ \ }S_{10}=(T_{1},-\mathbf{k}T_{1}).
\end{gather*}

Each of the ten sets gives a copy of a $\{3,4,3\}$ polytope on a fiber $%
S^{3} $. The points can be Hopf mapped, as described elsewhere\cite%
{mosseri-dandolof}, onto the base space $S^{4}$. The location of the mapped
point is intimately related to the entanglement of the corresponding
two-qubit state. Indeed, the Hopf map is simply described as a first map
which sends the pair $(q_{1},q_{2})$ onto the quaternion $Q=\overline{%
q_{1}q_{2}^{-1}}$ (which is sent to infinity if $q_{2}=0$), followed by an
inverse stereographic map which sends $Q$ to $S^{4}.$ A main result is that
the Hopf map is sensitive to entanglement: for separable states, $Q$ is
simply a complex number, not a generic quaternion; conversely, for maximally
entangled states (MES), the purely complex part of $Q$ vanishes. This
translates onto the base $S^{4}$ in the following way. Embed $S^{4}$ into $%
R^{5},$ with coordinates $\left\{ x_{l},l=0\cdots4\right\} $; then separable
states are such that $x_{3}=x_{4}=0$, and the $S^{2}$ sphere spanned by $%
\left\{ x_{0},x_{1},x_{2}\right\} $ form the standard Bloch sphere of the
first qubit. Maximally entangled\emph{\ }states map onto the unit circle in
the plane $\left( x_{3},x_{4}\right) $. Note that a well known entanglement
measure, the concurrence\cite{wootters}, is simply given by the radius in
the plane $\left( x_{3},x_{4}\right) $: $c=\sqrt{x_{3}^{2}+x_{4}^{2}}$, an
expression which will be used later.

In the present case, it is then easy to verify that the sets $S_{1}$ to $%
S_{6}$ correspond to separable states, while sets $S_{7}$ to $S_{10}$
correspond to maximally entangled states. The correspondence between
vertices and states is made by transforming back the quaternion pairs into
complex quadruplets whose terms are ($t_{00},t_{01},t_{10},t_{11}$). More
precisely, the $(q_{1},q_{2})$ pair reads $\left( t_{00}+t_{01}\mathbf{j,}%
\text{ }t_{10}+t_{11}\mathbf{j}\right) .$ Note that the quaternion unit $%
\mathbf{j}$ acts on the right of the complex numbers, while it acts on the
left in the definition of $S_{7,8}$. Since quaternion multiplication is
non-commutative, this distinction is important in going back and forth
between the lattice points and the states.

\subsubsection{The separable states}

Consider the set $S_{1},$ corresponding to the twenty-four states such that $%
t_{10}=t_{11}=0$, and which reads 
\begin{equation*}
\left\vert 0\right\rangle _{1}\otimes\left\vert \Phi_{l},\omega\right\rangle
_{2},\;\;\omega=0,\pi/2,\pi,3\pi/2,\text{ and \ \ }l=1..6.
\end{equation*}
As a whole, the six sets $S_{1}\cdots$ $S_{6}$ encompass $6\times24=144$
vertices, forming altogether $36$ physical states, with four values of the
global phase for each qubit state. Note that the precise value of the phases
are important here in order that our discretization procedure uniformly
cover the full Hilbert space. Using the above defined eigenstates of the
one-qubit Pauli matrices, these states read:%
\begin{equation*}
\begin{tabular}{|l|l|l|}
\hline
$\left\vert \pm x\right\rangle \otimes\left\vert \pm x\right\rangle
e^{i\left( \pi/4+m\pi/2\right) }$ & $\left\vert \pm x\right\rangle
\otimes\left\vert \pm y\right\rangle e^{i\left( \pi/4+m\pi/2\right) }$ & $%
\left\vert \pm x\right\rangle \otimes\left\vert \pm z\right\rangle
e^{im\pi/2}$ \\ \hline
$\left\vert \pm y\right\rangle \otimes\left\vert \pm x\right\rangle
e^{i\left( \pi/4+m\pi/2\right) }$ & $\left\vert \pm y\right\rangle
\otimes\left\vert \pm y\right\rangle e^{i\left( \pi/4+m\pi/2\right) }$ & $%
\left\vert \pm y\right\rangle \otimes\left\vert \pm z\right\rangle
e^{im\pi/2}$ \\ \hline
$\left\vert \pm z\right\rangle \otimes\left\vert \pm x\right\rangle
e^{im\pi/2}$ & $\left\vert \pm z\right\rangle \otimes\left\vert \pm
y\right\rangle e^{im\pi/2}$ & $\left\vert \pm z\right\rangle
\otimes\left\vert \pm z\right\rangle e^{i\left( \pi/4+m\pi/2\right) }$ \\ 
\hline
\end{tabular}
\ 
\end{equation*}
where $m=0,1,2,3$ \ triggers the global phase. Each of the nine entries
stands for the four possible sign combinations, leading to the announced
thirty-six physical states. A simple view of these separable states consists
in relating them to the \textquotedblleft product\textquotedblright\ of two
octahedra, each one belonging to the Bloch sphere of the individual qubits.

\subsubsection{The maximally entangled states}

The remaining four sets (altogether $4\times24=96$ sites) lead to a slightly
more subtle structure. We find a total of twenty-four different physical
MES, with four phase-distinct two-qubit states for each.\ But in the present
case, the phase-distinct states actually belong to two different\ sets,
either $\left( S_{7},S_{8}\right) $ or $\left( S_{9},S_{10}\right) $

As an example, we consider the set $S_{7}$ and enumerate the states
corresponding to, say, the quaternion pair $\left( \frac{1}{\sqrt{2}},\frac{1%
}{\sqrt{2}}\mathbf{j}\right) \in T_{1},$ which translates into the complex
quadruplet $(\frac{1}{\sqrt{2}},0,0,\frac{1}{\sqrt{2}})$ and therefore to
the MES: 
\begin{equation*}
\left\vert MES_{1},0\right\rangle =\frac{1}{\sqrt{2}}\left( \left\vert
00\right\rangle +\left\vert 11\right\rangle \right) =\frac{1}{\sqrt{2}}%
\left( \left\vert +z,+z\right\rangle +\left\vert -z,-z\right\rangle \right) .
\end{equation*}
The shortened notation $\left\vert +z,+z\right\rangle $ stands for $%
\left\vert +z\right\rangle _{1}\otimes\left\vert +z\right\rangle _{2}.$
There is only one other element in $S_{7},$ the pair $\left( -\frac{1}{\sqrt{%
2}},-\frac {1}{\sqrt{2}}\mathbf{j}\right) $, corresponding to $\left\vert
MES_{1},\pi\right\rangle =e^{i\pi}\left\vert MES_{1},0\right\rangle .$ The
other two elements belong to the set $S_{8}$ and read $\left( \frac{\mathbf{i%
}}{\sqrt{2}},\frac{\mathbf{k}}{\sqrt{2}}\right) $ and $\left( \frac {-%
\mathbf{i}}{\sqrt{2}},\frac{\mathbf{-k}}{\sqrt{2}}\right) ,$ associated
respectively to the states $\left\vert MES_{1},\pi/2\right\rangle $ and $%
\left\vert MES_{1},-\pi/2\right\rangle $. Before giving the full list of
states, it is interesting to focus on the next three states in $S_{7}$,
generated by $\mathbf{i,}$ $\mathbf{j}$ and $\mathbf{k}$. One gets,
respectively, 
\begin{align*}
\left\vert MES_{2},0\right\rangle & =\frac{\mathbf{i}}{\sqrt{2}}\left(
\left\vert 00\right\rangle -\left\vert 11\right\rangle \right) =\frac {%
\mathbf{i}}{\sqrt{2}}\left( \left\vert +z,+z\right\rangle -\left\vert
-z,-z\right\rangle \right) , \\
\left\vert MES_{3},0\right\rangle & =\frac{1}{\sqrt{2}}\left( \left\vert
01\right\rangle +\left\vert 10\right\rangle \right) =\frac{1}{\sqrt{2}}%
\left( \left\vert +z,-z\right\rangle +\left\vert -z,z\right\rangle \right) ,
\\
\left\vert MES_{4},0\right\rangle & =\frac{\mathbf{i}}{\sqrt{2}}\left(
\left\vert 01\right\rangle -\left\vert 10\right\rangle \right) =\frac {%
\mathbf{i}}{\sqrt{2}}\left( \left\vert +z,-z\right\rangle -\left\vert
-z,z\right\rangle \right) .
\end{align*}
The set $\left\{ \left\vert MES_{1},0\right\rangle ,\left\vert
MES_{2},0\right\rangle ,\left\vert MES_{3},0\right\rangle ,\left\vert
MES_{4},0\right\rangle \right\} $ forms the entangled \textquotedblleft
magic\textquotedblright\ basis described elsewhere\cite%
{bennett-96,hill-wootters}.\ Each physical state $\left\vert
MES_{l}\right\rangle $ corresponds to the four two-qubit states $\left\vert
MES_{l},\omega\right\rangle ,$ with $\omega=0,\pi$ for states in $S_{7}$ and 
$\omega=\pi/2,-\pi/2$ for states in $S_{8}.$ As a whole, the forty-eight
vertices of the sets $S_{7}$ and $S_{8}$ provide twelve physical MES, each
one representing a set of four phase-related states. They are listed below,
in the $\omega=0$ case, and with $\epsilon=e^{i\pi/4}$: 
\begin{align*}
\left\vert MES_{5},0\right\rangle & =\frac{\epsilon}{2}\left( \left\vert
00\right\rangle +\left\vert 01\right\rangle \right) -\frac{\overline {%
\epsilon}}{2}\left( \left\vert 10\right\rangle -\left\vert 01\right\rangle
\right) =\frac{\epsilon}{\sqrt{2}}\left( \left\vert +z,+x\right\rangle +%
\mathbf{i\,}\left\vert -z,-x\right\rangle \right) , \\
\left\vert MES_{6},0\right\rangle & =\frac{\epsilon}{2}\left( \left\vert
00\right\rangle -\left\vert 10\right\rangle \right) +\frac{\overline {%
\epsilon}}{2}\left( \left\vert 01\right\rangle +\left\vert 11\right\rangle
\right) =\frac{\epsilon}{\sqrt{2}}\left( \left\vert z,-y\right\rangle
-\left\vert -z,y\right\rangle \right) , \\
\left\vert MES_{7},0\right\rangle & =\frac{\overline{\epsilon}}{2}\left(
\left\vert 00\right\rangle -\left\vert 10\right\rangle \right) +\frac {%
\epsilon}{2}\left( \left\vert 01\right\rangle +\left\vert 11\right\rangle
\right) =\frac{\overline{\epsilon}}{\sqrt{2}}\left( \left\vert
+z,+y\right\rangle -\left\vert -z,-y\right\rangle \right) , \\
\left\vert MES_{8},0\right\rangle & =\frac{\overline{\epsilon}}{2}\left(
\left\vert 00\right\rangle +\left\vert 01\right\rangle \right) -\frac {%
\epsilon}{2}\left( \left\vert 10\right\rangle -\left\vert 11\right\rangle
\right) =\frac{\overline{\epsilon}}{\sqrt{2}}\left( \left\vert
+z,+x\right\rangle -\mathbf{i\,}\left\vert -z,-x\right\rangle \right) , \\
\left\vert MES_{9},0\right\rangle & =\frac{\epsilon}{2}\left( \left\vert
00\right\rangle -\left\vert 01\right\rangle \right) +\frac{\overline {%
\epsilon}}{2}\left( \left\vert 10\right\rangle +\left\vert 11\right\rangle
\right) =\frac{\epsilon}{\sqrt{2}}\left( \left\vert +z,-x\right\rangle -%
\mathbf{i\,}\left\vert -z,x\right\rangle \right) , \\
\left\vert MES_{10},0\right\rangle & =\frac{\epsilon}{2}\left( \left\vert
00\right\rangle +\left\vert 10\right\rangle \right) -\frac{\overline {%
\epsilon}}{2}\left( \left\vert 01\right\rangle -\left\vert 11\right\rangle
\right) =\frac{\epsilon}{\sqrt{2}}\left( \left\vert +z,+y\right\rangle
+\left\vert -z,-y\right\rangle \right) , \\
\left\vert MES_{11},0\right\rangle & =\frac{\overline{\epsilon}}{2}\left(
\left\vert 00\right\rangle +\left\vert 10\right\rangle \right) -\frac {%
\epsilon}{2}\left( \left\vert 01\right\rangle -\left\vert 11\right\rangle
\right) =\frac{\overline{\epsilon}}{\sqrt{2}}\left( \left\vert
+z,-y\right\rangle \mathbf{+}\left\vert -z,y\right\rangle \right) , \\
\left\vert MES_{12},0\right\rangle & =\frac{\overline{\epsilon}}{2}\left(
\left\vert 00\right\rangle -\left\vert 01\right\rangle \right) +\frac {%
\epsilon}{2}\left( \left\vert 10\right\rangle +\left\vert 11\right\rangle
\right) =\frac{\overline{\epsilon}}{\sqrt{2}}\left( \left\vert
+z,-x\right\rangle +\mathbf{i\,}\left\vert -z,x\right\rangle \right) .
\end{align*}
The latter twelve physical states may be read off from the two sets $S_{9}$
and $\ S_{10}$:%
\begin{align*}
\left\vert MES_{13},0\right\rangle & =\frac{1}{\sqrt{2}}\left\vert
00\right\rangle +\frac{\mathbf{i}}{\sqrt{2}}\left\vert 11\right\rangle =%
\frac{1}{\sqrt{2}}\left( \left\vert +z,+z\right\rangle +\mathbf{i}\text{\/}%
\left\vert -z,-z\right\rangle \right) , \\
\left\vert MES_{14},0\right\rangle & =\frac{\mathbf{i}}{\sqrt{2}}\left\vert
00\right\rangle +\frac{1}{\sqrt{2}}\left\vert 11\right\rangle =\frac{1}{%
\sqrt{2}}\left( \mathbf{i\,}\left\vert +z,+z\right\rangle +\text{\/}%
\left\vert -z,-z\right\rangle \right) , \\
\left\vert MES_{15},0\right\rangle & =\frac{1}{\sqrt{2}}\left\vert
01\right\rangle -\frac{\mathbf{i}}{\sqrt{2}}\left\vert 10\right\rangle =%
\frac{1}{\sqrt{2}}\left( \left\vert +z,-z\right\rangle -\mathbf{i}\text{\/}%
\left\vert -z,+z\right\rangle \right) , \\
\left\vert MES_{16},0\right\rangle & =\frac{\mathbf{i}}{\sqrt{2}}\left\vert
01\right\rangle -\frac{1}{\sqrt{2}}\left\vert 10\right\rangle =\frac{1}{%
\sqrt{2}}\left( \mathbf{i\,}\left\vert +z,-z\right\rangle -\left\vert
-z,+z\right\rangle \right) , \\
\left\vert MES_{17},0\right\rangle & =\frac{\epsilon}{2}\left( \left\vert
00\right\rangle +\left\vert 01\right\rangle \right) -\frac{\mathbf{i}%
\overline{\epsilon}}{2}\left( \left\vert 10\right\rangle -\left\vert
11\right\rangle \right) =\frac{\epsilon}{\sqrt{2}}\left( \left\vert
+z,+x\right\rangle -\left\vert -z,-x\right\rangle \right) , \\
\left\vert MES_{18},0\right\rangle & =\frac{\epsilon}{2}\left( \left\vert
00\right\rangle -\mathbf{i\,}\left\vert 10\right\rangle \right) +\frac{%
\overline{\epsilon}}{2}\left( \left\vert 01\right\rangle +\mathbf{i\,}%
\left\vert 11\right\rangle \right) =\frac{\epsilon}{\sqrt{2}}\left(
\left\vert +z,-y\right\rangle -\mathbf{i\,}\left\vert -z,+y\right\rangle
\right) , \\
\left\vert MES_{19},0\right\rangle & =\frac{\overline{\epsilon}}{2}\left(
\left\vert 00\right\rangle -\mathbf{i\,}\left\vert 10\right\rangle \right) +%
\frac{\epsilon}{2}\left( \left\vert 01\right\rangle +\mathbf{i\,}\left\vert
11\right\rangle \right) =\frac{\overline{\epsilon}}{\sqrt{2}}\left(
\left\vert +z,+y\right\rangle -\mathbf{i\,}\left\vert -z,-y\right\rangle
\right) , \\
\left\vert MES_{20},0\right\rangle & =\frac{\overline{\epsilon}}{2}\left(
\left\vert 00\right\rangle +\left\vert 01\right\rangle \right) -\frac {%
\mathbf{i}\epsilon}{2}\left( \left\vert 10\right\rangle -\left\vert
11\right\rangle \right) =\frac{\overline{\epsilon}}{\sqrt{2}}\left(
\left\vert +z,+x\right\rangle +\left\vert -z,-x\right\rangle \right) , \\
\left\vert MES_{21},0\right\rangle & =\frac{\epsilon}{2}\left( \left\vert
00\right\rangle -\left\vert 01\right\rangle \right) +\frac{\mathbf{i}%
\overline{\epsilon}}{2}\left( \left\vert 10\right\rangle +\left\vert
11\right\rangle \right) =\frac{\epsilon}{\sqrt{2}}\left( \left\vert
+z,-x\right\rangle +\left\vert -z,x\right\rangle \right) , \\
\left\vert MES_{22},0\right\rangle & =\frac{\epsilon}{2}\left( \left\vert
00\right\rangle +\mathbf{i\,}\left\vert 10\right\rangle \right) -\frac{%
\overline{\epsilon}}{2}\left( \left\vert 01\right\rangle -\mathbf{i\,}%
\left\vert 11\right\rangle \right) =\frac{\epsilon}{\sqrt{2}}\left(
\left\vert +z,+y\right\rangle +\mathbf{i\,}\left\vert -z,-y\right\rangle
\right) , \\
\left\vert MES_{23},0\right\rangle & =\frac{\overline{\epsilon}}{2}\left(
\left\vert 00\right\rangle +\mathbf{i\,}\left\vert 10\right\rangle \right) -%
\frac{\epsilon}{2}\left( \left\vert 01\right\rangle -\mathbf{i\,}\left\vert
11\right\rangle \right) =\frac{\overline{\epsilon}}{\sqrt{2}}\left(
\left\vert +z,-y\right\rangle +\mathbf{i\,}\left\vert -z,y\right\rangle
\right) , \\
\left\vert MES_{24},0\right\rangle & =\frac{\overline{\epsilon}}{2}\left(
\left\vert 00\right\rangle -\left\vert 01\right\rangle \right) +\frac {%
\mathbf{i}\epsilon}{2}\left( \left\vert 10\right\rangle +\left\vert
11\right\rangle \right) =\frac{\overline{\epsilon}}{\sqrt{2}}\left(
\left\vert +z,-x\right\rangle -\left\vert -z,x\right\rangle \right) .
\end{align*}

Succinctly, the ninety-six two-qubit entangled states can be written: 
\begin{equation*}
\left\vert MES_{l},\omega\right\rangle =e^{i\omega}\left\vert
MES_{l},0\right\rangle ,\text{ \ with \ \ }\omega=0,\pi/2,\pi,3\pi/2\text{ \
\ and \ }l=1\cdots24.
\end{equation*}
Modding out the global phase, we can write the twenty-four physical states
in the form:%
\begin{equation*}
\begin{tabular}{|l|l|}
\hline
$\frac{1}{\sqrt{2}}\left( \left\vert +z,+z\right\rangle +e^{i\theta
}\left\vert -z,-z\right\rangle \right) $ & $\frac{1}{\sqrt{2}}\left(
\left\vert +z,-z\right\rangle +e^{i\theta}\left\vert -z,z\right\rangle
\right) $ \\ \hline
$\frac{1}{\sqrt{2}}\left( \left\vert +z,+x\right\rangle +e^{i\theta
}\left\vert -z,-x\right\rangle \right) $ & $\frac{1}{\sqrt{2}}\left(
\left\vert +z,-x\right\rangle +e^{i\theta}\left\vert -z,+x\right\rangle
\right) $ \\ \hline
$\frac{1}{\sqrt{2}}\left( \left\vert +z,+y\right\rangle +e^{i\theta
}\left\vert -z,-y\right\rangle \right) $ & $\frac{1}{\sqrt{2}}\left(
\left\vert +z,-y\right\rangle +e^{i\theta}\left\vert -z,+y\right\rangle
\right) $ \\ \hline
\end{tabular}
\ 
\end{equation*}
with $\theta=0,\pi/2,\pi,3\pi/2$.

Note that these twenty-four entangled states, together with the above
thirty-six separable states, are in one-to-one correspondence, up to a
global phase, with the sixty discrete states on $\mathfrak{H}_{2}$ presented
in section 2.

\subsection{Comments}

\subsubsection{Finer discretizations of $H_{2}$: higher $E_{8}$ shells}

Thus far, this alternate technique has provided compelling confirmation of
the results from section 2. However, the present purely geometric approach
does not describe the discrete set's transformation group. This point could
be in principle addressed, albeit in a much less transparent way than in the
first part of this paper, by using the $E_{8}$ lattice point group (which
has a high order) and finding the subgroup of rotations which survive the
Hopf map. But this lattice approach does have the benefit of allowing
discrete sets with a finer minimum distance (i.e. $0<d_{jk}<\pi/2$) to be
explored in a straightforward manner: while the first shell of $E_{8}$
provides the discrete set with a minimum distance of $\pi/2$, a finer
discretization might be achieved by considering the higher order shells in $%
E_{8}$. This construction would provide a uniform set of two-qubit states,
some of which would have intermediate entanglement. A note of caution is in
order here, since we are only interested in describing normalized quantum
states. Lattice points which are aligned, as viewed from the origin,
contribute to the same two-qubit state. One should therefore only focus on
the \textquotedblleft visible points,\textquotedblright\ which form the
lattice's M\"{o}bius transform\cite{mosseri-visible}.

We do not give here a detailed description of these finer discretizations of
\thinspace$H_{2}$. However, we note that the number $M_{J}$ of sites on the $%
J^{th}$ shell around an $E_{8}$ vertex is simply given by\cite{conway-sloane}
\begin{equation*}
M_{J}=240\sum_{d\mid J}d^{3},
\end{equation*}
where $d$ denotes integers which divide $J$. The table below displays these
numbers for the first four shells. Again, the physical states are obtained
from these two-qubit states by modding out a global phase.

\begin{center}
\begin{tabular}{|c|c|c|c|c|}
\hline
$J$ & $1$ & $2$ & $3$ & $4$ \\ \hline
$M$ & $240$ & $2160$ & $6720$ & $17520$ \\ \hline
\end{tabular}
\end{center}

The shell by shell analysis, and its relation to the Hopf map, was done
elsewhere\cite{sadoc-mosseri-livre,sadoc-mosseri-E8}. \ It allows us to get
points on the second shell corresponding to states having concurrence $%
0,1/2,1/\sqrt{2},1$.\ The third \ shell contributes states of concurrence $%
0, $ $1/3,$ $2/3,$ $\sqrt{5}/3,$ $\sqrt{8}/3$ and $1.$

\subsubsection{Discrete one-qubit \textit{mixed} states}

A second advantage of our lattice approach over the pseudostabilizer
approach is that it allows a discussion of discrete sets of \textit{mixed}
states. It is well known that the full set of one-qubit mixed states can be
obtained by tracing out one qubit of generic two-qubit pure states. Mixed
states are represented by points inside the so-called \textquotedblleft
Bloch ball,\textquotedblright\ bounded by the pure-state Bloch sphere. In
the context of generalizing the Bloch sphere for two-qubit pure states using
the $S^{7}$ Hopf fibration\cite{mosseri-dandolof}, it was shown that the
Bloch ball corresponds precisely to the restriction to the triplet $\left(
x_{0},x_{1},x_{2}\right) $ on the base space $S^{4}$. This describes mixed
states obtained upon tracing out the second qubit. With this in mind, we are
tempted to propose, in parallel to the two-qubit pure state discretization,
an $E_{8}$-related discretization of the Bloch ball.

From the $E_{8}$ first shell, one gets the six permutations $\left(
\pm1,0,0\right) $, corresponding to pure states on the Bloch sphere (the
traced separable two-qubit states) forming an octahedron. But one also gets
the point $\left( 0,0,0\right) $, the Bloch ball centre, corresponding to
traced maximally entangled states. Then, from the $E_{8}$ second shell, we
find in addition the eight permutations $\left( \pm%
{\frac12}%
,\pm%
{\frac12}%
,\pm%
{\frac12}%
\right) $---a cube of radius $\sqrt{3}/2$---and the twelve permutations $%
\left( \pm%
{\frac12}%
,\pm%
{\frac12}%
,0\right) $---a cuboctahedron of radius $1/\sqrt{2}.$ The Bloch ball can be
further discretized by using traced states originating from higher $E_{8}$
shells.

\subsubsection{Conclusion to section 3}

We have presented a second technique for producing a uniform discrete set of
states from the continuous Hilbert space, again focusing on the one- and
two-qubit cases. The pseudostabilizer and dense lattice strategies were done
independently, and their agreement provides a compelling confirmation of our
results, while also calling for a better understanding of their possible
deeper relationship. While the dense lattice strategy is less easily focused
on the transformation groups leaving the discrete sets invariant, it does
have the advantage of allowing a discussion of discrete sets with a smaller
minimum distance between states, and also of discrete sets of mixed states.
Looking ahead, a discretization of higher-dimensional Hilbert spaces may be
achieved in the dense lattice approach by using high-dimensional lattices in 
$2^{N}$ dimensions\cite{elser-sloane}. Though it does not describe the
entanglement properties as nicely as in the two-qubit case, the $S^{15}$
Hopf map, corresponding to the three-qubit case, has been found to be
entanglement sensitive\cite{mosseri-dresde,bernevig-chen}. This case should
be related to the dense lattice $\Lambda_{16}$ in $%
\mathbb{R}
^{16}$, and is presently under study. It is interesting to note that the
number of lattice sites closest to the origin---the lattice
\textquotedblleft kissing number\textquotedblright---for this case is 4320,
which is precisely four times the expected number of vertices on the uniform
Hilbertian polytope $\mathfrak{H}_{3}$. We are therefore likely to face a
similar situation as in the one- and two-qubit cases, where there were four
phase-related qubit states associated with each physical state. However, the
four-to-one relation between the $\Lambda_{16}$ first shell sites and the
vertices of $\mathfrak{H}_{3}$ remains to be checked. Generalization to more
than three qubits cannot use the Hopf fibrations, limited to $S^{15}$. A
particularly interesting family to be checked further is the one described
long ago by John Leech\cite{leech}, which coincides with those studied here
for $N=1,2$ and $3$, and whose kissing number is, for any $N$, precisely
four times that given in the first part of this paper for the number of
states in the generic Hilbertian polytopes. As for the three-qubit case,
this precise four-to-one relation remains to be checked.

\bigskip

\textit{M.D. and C.R. are indebted to Ike Chuang for his enthusiastic
teaching of quantum information theory and to Daniel Gottesman for
communicating and explaining his results. In addition, R.M. would like to
thank Perola Milman and Karol Zyczkowski for interesting discussions and
comments on the two-qubit Hilbert space geometry, and Philippe Biane who,
while reading the first version of this manuscript, drew his attention to
John Leech's 1964 paper. Daniel Esteve and Denis Vion are also gratefully
acknowledged by all of us for helpful interactions. Finally, John Preskill's
web-accessible notes and exercises on quantum information have been very
useful. This work has been supported by the ARO/ARDA grant
DAAD190210044.\bigskip}

\end{document}